\documentclass[11pt,preprint]{aastex}

\begin{document}
 
\title{An Infrared Coronagraphic Survey for Substellar Companions}

\author{Patrick J. Lowrance\altaffilmark{1}, E. E. Becklin\altaffilmark{2},
Glenn Schneider\altaffilmark{3}, J. Davy Kirkpatrick\altaffilmark{4}, 
Alycia J. Weinberger\altaffilmark{5}, B. Zuckerman\altaffilmark{2}, 
Christophe Dumas\altaffilmark{6}, Jean-Luc Beuzit\altaffilmark{7},    
Phil Plait\altaffilmark{8}, Eliot Malumuth\altaffilmark{9}, 
Sally Heap\altaffilmark{9}, 
Richard J. Terrile\altaffilmark{10}, Dean C. Hines\altaffilmark{11}}

\altaffiltext{1}{Spitzer Science Center, MS 220-6, California Institute of Technology, Pasadena, CA. 91125; lowrance@ipac.caltech.edu}

\altaffiltext{2}{University of California Los Angeles, Los Angeles, CA}

\altaffiltext{3}{University of Arizona, Tucson, AZ}

\altaffiltext{4}{Infrared Processing and Analysis Center, California
Institute of Technology, Pasadena, CA}

\altaffiltext{5}{Carnegie Institution of Washington, Department of Terrestial Magnetism, Washington, DC}

\altaffiltext{6}{European Southern Observatory, Santiago, Chile}

\altaffiltext{7}{Laboratoire d'Astrophysique, Observatoire de Grenoble, France}

\altaffiltext{8}{Sonoma State University, Rohnert Park, CA}

\altaffiltext{9}{NASA, Goddard Space Flight Center, Greenbelt, MD}

\altaffiltext{10}{Jet Propulsion Laboratory, Pasadena, CA }

\altaffiltext{11}{Space Science Institute, Boulder, CO}

\begin{abstract}
We have
used the  F160W filter (1.4 $-$ 1.8 $\mu$m) and the coronagraph on the 
Near-InfraRed Camera and Multi-Object Spectrometer (NICMOS) on the
Hubble Space Telescope (HST) to survey 45 single stars with a
median age of 0.15 Gyr, an average distance of 30 pc, and an average
H-magnitude of 7 mag. For the median age 
we were capable of detecting a 30 M$_{Jupiter}$ companion 
at separations between 15 and 200 AU. A 5 M$_{Jupiter}$ object could
have been detected at 30 AU around 36\% of our primaries. 
For several of our targets that were less than 30 Myr old,
the lower mass limit was as low as a Jupiter mass, well into the
high mass planet region. Results of the entire survey 
include the proper motion verification of 
five low-mass stellar companions, two brown dwarfs (HR7329B and TWA5B)
and one possible brown dwarf binary (Gl 577B/C). 
\end{abstract}


\section{Introduction}

Just over a decade ago substellar astronomy exploded in discoveries 
with the first brown dwarfs in the Pleiades (Rebolo, Zapatero Osorio \& Martin 1995), 
the first `cool' brown dwarf, Gl 229B (Nakajima et al. 1995), and 
the first extrasolar planet (Mayor \& Queloz 1995).
The added identification of hundreds of field brown dwarfs in large scale surveys such 
as 2MASS (Skrutskie et al 1997) and SDSS (York et al 2000) led to the 
creation of two new spectral types, L and T (Kirkpatrick et al 1999; 
Burgasser et al 2002; Geballe et al 2002). 

Brown dwarfs occupy the important region in the mass range 
between stars and planets. Their existence and  
properties give insight into stellar and planetary formation. 
They are thought to form like stars, but do
not have enough mass to sustain hydrogen fusion. With low temperatures and compact atmospheres 
different from stars, the lowest-mass brown dwarfs should 
resemble the higher mass giant planets. The observational distinctions between 
a planet and brown dwarf at the low mass end have yet to be constructed except that 
a 'planet' should be a companion to a star.   

To fully explore the 
similarities and differences of stellar and planetary formation, 
brown dwarfs as companions are essential. Since the 
brighter primary's age and distance are known, these two properties 
that are usually uncertain for field brown dwarfs are already established for companions. 
Therefore companions can be standards and anchor models as 
the multiplicity fraction, separation distribution, and mass ratio 
distribution of companions provide important hints as to formation mechanisms.
Radial velocity surveys find that about 5\% of G stars have 
massive Jupiter planets within 5 AU, but that brown dwarfs are rare ($<$0.1\%) 
at these separations. Even though Gl 229B is a companion to an M star, 
many earlier companion searches have turned up few discoveries.
Closer looks at the younger cluster and field 
brown dwarfs have found several to be equal-mass binaries 
(Zapatero-Osorio, Martin,\& Rebolo
1997; Koerner et al 1999; Reid et al 2001). About 
20\% of field L dwarfs have been shown to be binaries with projected 
separations $<$ 20AU while closer to 45\% of L dwarf companions to stars 
are actually binaries (Burgasser, Kirkpatrick, \& Lowrance 2005). 
The frequency of brown dwarf companions might be
dependent on mass and separation (Gizis et al. 2001), but these ideas need a larger 
sample to be explored.

The primary goal of this survey was the discovery of
substellar companions to young, nearby stars. 
Here we present results from our infrared coronagraphic survey with the 
Near-Infrared Camera and Multi-Object Spectrometer (NICMOS) on the Hubble 
Space Telescope (HST), including follow-up adaptive 
optics observations obtained 
at the Canada-France-Hawaii Telescope (CFHT) and 200in Hale Telescope 
and spectra obtained with the Space Telescope Imaging Spectrograph (STIS).
Previously, this survey presented TWA5B, a $\sim$ 20 Jupiter mass brown
dwarf companion (Lowrance et al. 1999) and HR 7329B, a $\sim$ 40 Jupiter mass brown
dwarf companion (Lowrance et al. 2000). The full survey results along with six additional 
discoveries are presented here.

\section{Defining the Sample}

 As part of the ISO Debris project to detect dust around 
main sequence stars, a list of young, nearby stars was assembled. 
The list was expanded to include young objects selected 
through techniques including chromospheric
activity, the presence of lithium, and space 
motions as discussed later. 
Typically, the systems selected were young, single stars within 
$\sim$ 50 pc of the Sun. Since brown dwarfs are relatively dim, younger and
closer ones should be brighter, and therefore easier to detect. 

Brown dwarfs and giant planets are more luminous when they are young 
due to gravitational contraction energy remaining from formation. Models 
of substellar objects show that $L \sim t^{-1.3} 
M^{2.24}$ , where L is luminosity, t is age and M is mass, 
based on cooling curves using atmospheric physics (Burrows et al. 1997). 
Since models predict the bolometric correction 
at H-band does not
change much for cool temperatures below 3000 K, we can 
expect a similar relation for an object's H-band flux 
as a function of mass and age. 
If atmospheric models are correct, we expect a 5 M$_{Jupiter} $\ 
object to have a M$_H \sim 17.5$ mag at $10^8$ years (Burrows et al. 1997).

\subsection{Building the Target List}

As part of the NICMOS Instrument Definition Team (IDT), 
the EONS team (Environments of Nearby
Stars) was awarded guaranteed time of approximately 80 orbits on the
Hubble Space Telescope, of which 50 were dedicated to 
finding brown dwarfs around young stars (HST Program ID 7226, 7227). 
The team prioritized the list of stars 
to the "fifty best" targets, attempting to maximize the
probability of discovery. 
The three main factors that were taken into 
account include lack of binarity, close distance, and young age. 

A close ($<$20$\arcsec$) binary significantly 
lowers the probability of another companion being in a 
close, stable orbit (within a few arcsec) 
to the star. Triple systems with a ratio of major axes $<$ 5 are
thought to be unstable (Harrington 1977). 
All known close binaries were left off the target list,
but since many of the apparently single stars had never been observed with
at high resolution, some stellar binaries were still to be expected. 

At a distance of 50 parsecs, 
objects discovered at 0$\farcs$4 (coronagraphic radius $=$ 0$\farcs$3)
would be 20 AU from the primary
star. This is near the maximum in distribution of separations of binaries in
Duquennoy \& Mayor (1991). 
Most of the distances for the targets are determined 
from the Hipparcos mission data,
and are accurate to a few percent. Some of the stars were not observed
by the Hipparcos mission and the photometric distance is used. The 
distribution of distances peaks near 30 parsecs, the median
distance of the sample. The minimum observable 
separation of 12 AU, at 30 pc, is inside the average orbital distance of 
giant planets in our solar system. A few stars selected for 
this project were farther than 50 parsecs but were choosen because of their 
extreme youth (e.g., TW Hydrae Association).

There are many stars within $\sim$ 50 pc of the sun with ages 
similar to or less than the age of the Pleiades 
(t$=$125 Myr, Stauffer, Schultz, \& Kirkpatrick 1998). 
Youth can be inferred observationally in a number of ways. These different 
age determinations can be intercompared. 
One method is measuring photospheric lithium
abundance in stars later than $\sim$K5 in spectral type. 
This fragile element is destroyed at temperatures greater 
than 2 $\times$ 10$^6$K and convective motions within a late-type 
star's layers will, within less than a hundred million years for 0.1M$_{\sun}$ stars, 
cycle all lithium to the hot core. Studies of stars in clusters (Favata et al. 1993) 
have shown that at a given spectral type, lithium
abundance decreases as a function of age, where age can be determined
independently from other means, such as main-sequence fitting.

Another method to determine age is measurement of coronal or chromospheric activity such as 
H$\alpha$ emission, Ca H \& K line emission, and X-ray emission. All
of these are coupled to rotational velocity and 
magnetic field activity, presumably a result of the internal dynamo. 
Stars typically spin up to high
rotational velocities ($>200$km s$^{-1}$) when young and slow down 
as angular momentum is lost through stellar winds. 
A problem with this method on an individual star basis 
is that close binaries with periods less
than a few weeks are often tidally locked and maintain high levels of
activity regardless of age. 

Main-seqeunce fitting is another useful method 
which places stars on an H-R diagram and compares their location 
to clusters of approximate ages. Accurate distances of field stars became 
available as a result of the Hipparcos mission, and 
main-sequence fitting became much more effective in the age determination.
One method of age determination 
is kinematics, in which comparisons
are made between a star's galactic motion and that of clusters 
of approximate ages. This
proves successful in providing a general age-group for individual
stars. 

Age estimates for our our sample are found using an 
intercomparison of lithium abundance, coronal \& chromospheric activity, 
rotational velocity, main-sequence fitting, and kinematics (Appendix A). 
Upper limits and lower limits are calibrated against 
samples of T Tauri stars ($\sim$1 Myr) and 
stars in young, nearby clusters such as 
the Pleiades (125 Myr), Ursa Major (300 Myr) and the Hyades (600
Myr).

The original plan was to 
re-observe all frames containing point-sources within three to four
years for proper motion confirmation of companionship. 
However, it was learned after launch that NICMOS would have a shorter
lifetime than expected due to coolant problems. Chance for
follow-up was now limited to within two-years with NICMOS
or from ground based telescopes. 
To limit the number of extraneous background stars, 
we therefore excluded most 
target stars with $\mid b \mid < 15\deg$.

\begin{figure}
\figurenum{1}
\epsscale{1.0}
\plottwo{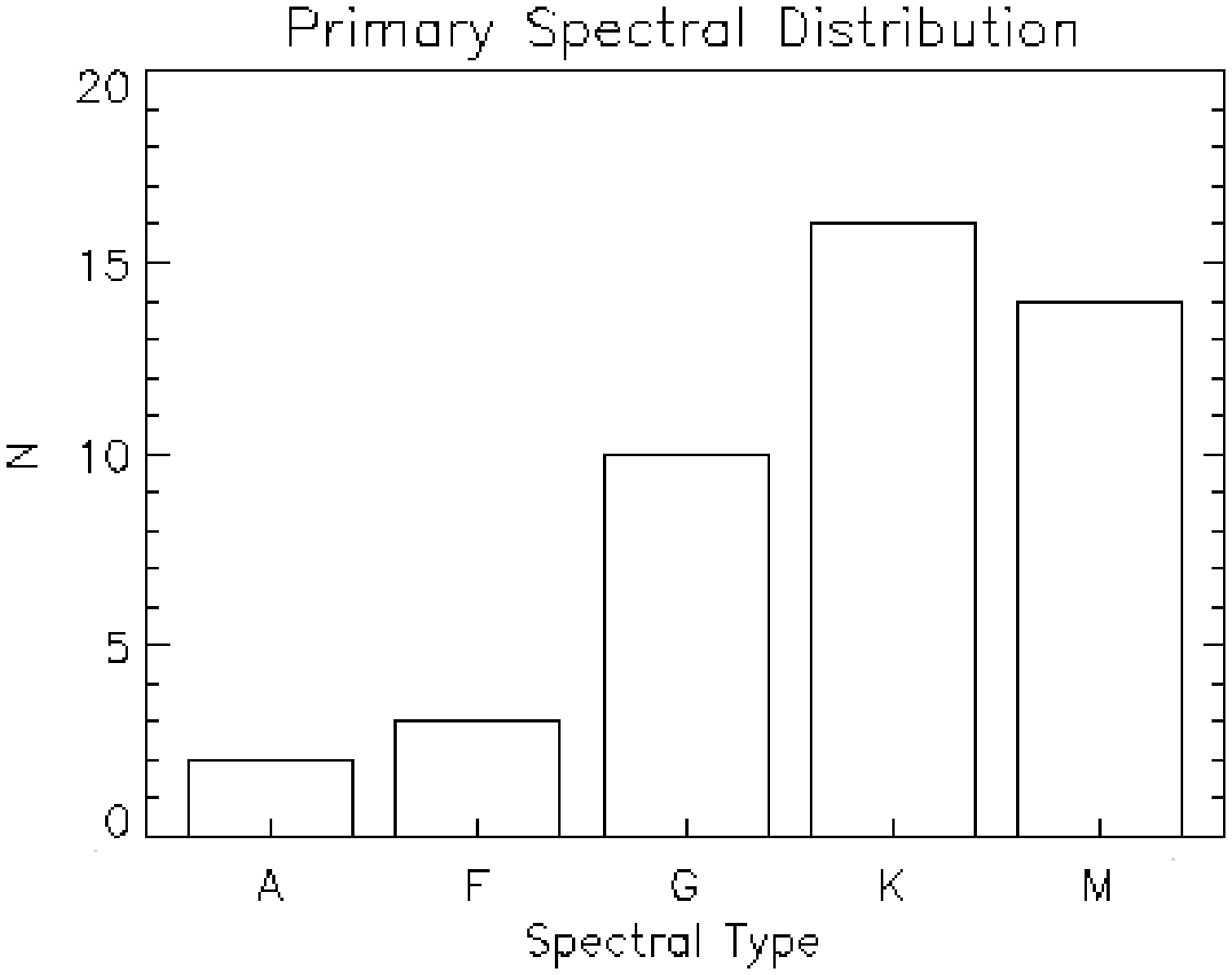}{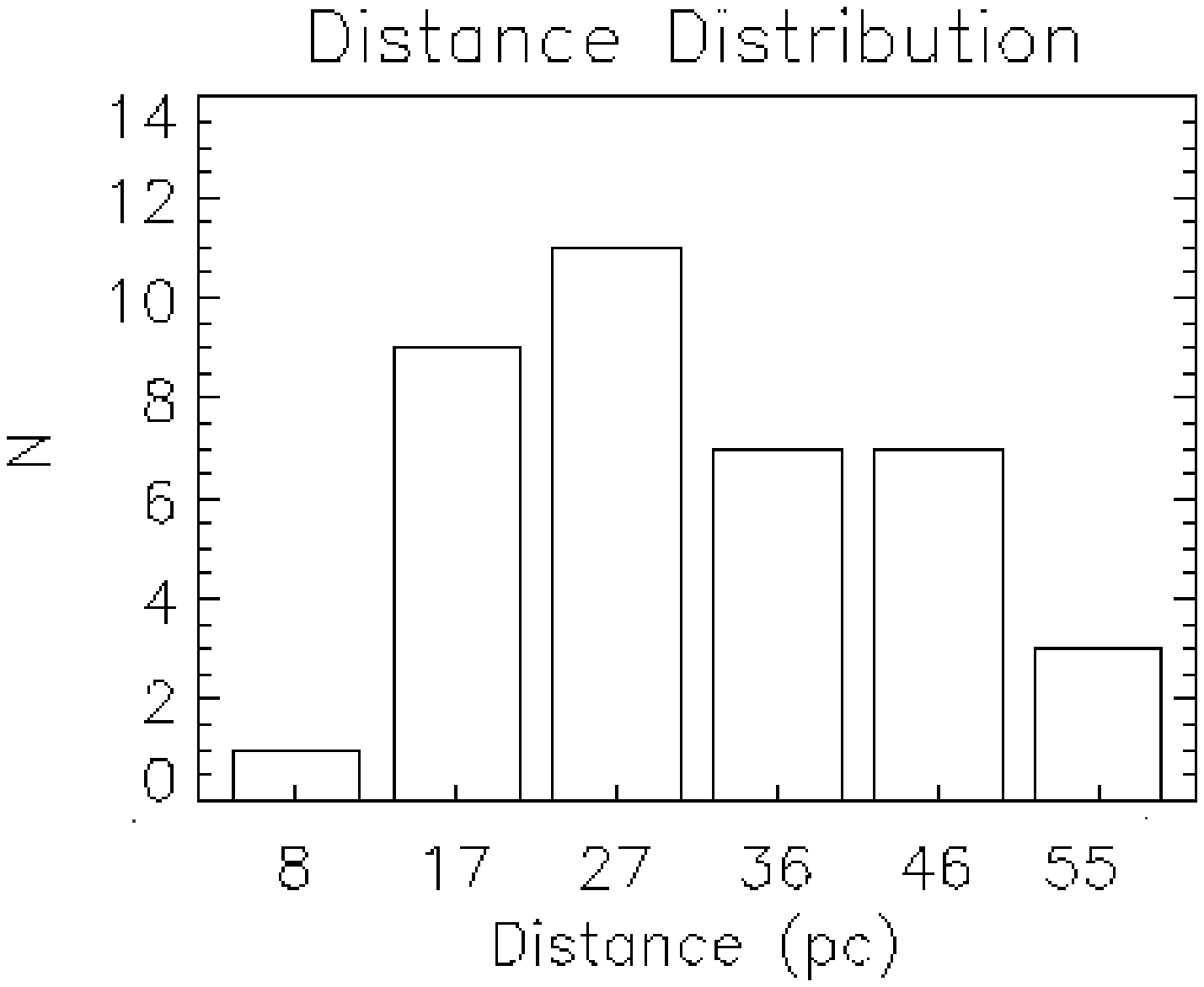}
\figcaption{The sample consists of a range of spectral types and
distances, with a concentration toward lower mass ($<$1.0 M${_\sun}$) stars 
at a median distance of about 30 parsecs. Half of the distances, or 25, 
were determined by
data from the Hipparcos mission.}\label{barplts}
\end{figure}

The final sample (Table 1) consists of 45 single stars with 
a median distance of
30 parsec and median age of 0.15 Gyr. 
The sample was concentrated toward cooler
spectral types  (Figure \ref{barplts}) mainly for 
contrast considerations, because faint
substellar companions will be easier to detect near fainter
primaries.  They were observed as scheduled from 18 Mar 1998 to 
18 Dec 1998.

\section{Observations}

\subsection{Using NICMOS: An Infrared Instrument Aboard the Hubble Space Telescope}
		
The main problem with trying to image brown dwarfs or giant planets around
main-sequence stars is the
overwhelming brightness of the primary. A substellar companion will
be much fainter than the star it orbits (i.e. a Gl
229B-like object, L$=$2$\times10^{-5}$L$_{\odot}$, orbiting a
solar-like star of 1 L$_{\odot}$). The cool 
brown dwarf makes up a little of this in the infrared, where it radiates much 
of its light, and is brighter with youth, but the primary is still
much brighter. According to theory (Burrows et al. 1997), 
Gl 229B was 50 times, or 4.25 mag, brighter at 10$^8$ yr old. 
Also, formation theories of high mass planets predicted orbits
very close to its primary, as well, so high resolution, 
achievable from space, was essential.
The Near-Infrared Camera and Multi-Object Spectrometer, 
NICMOS, is a second-generation instrument for 
Hubble Space Telescope. NICMOS has three available cameras. 
Camera 2 has a $\sim19\arcsec\times$19$\arcsec$ field of 
view with 0$\farcs$076 pixels and provides higher resolution 
and image stability then ground based instruments. 
Most importantly, the coronagraph on camera 2 is actually a hole 
in the field divider mirror that is baffled with a cold pupil 
plane mask (Thompson et al. 1998), and 
therefore can provide higher sensitivity to faint companions 
than direct imaging (and PSF subtraction) because of the actual 
reduction of light in the optical path. 

We conducted the survey at F160W, 
which corresponds roughly to the H-band filter on the ground. 
There are three main reasons for choosing this 
filter. First, the dominant sources of background radiation 
are the zodiacal 
light at short wavelengths and the thermal background emission from 
the telescope at long wavelengths. The sum of these two components are 
a minimum at 1.6$\mu$m (see Table 4.8 of the NICMOS Data Handbook). 
Second, in camera 2 the PSF is Nyquist
sampled at 1.67$\mu$m and $\sim$90\% of the light of the primary is
contained within the coronagraph which has a 0$\farcs$6 diameter. 
Because the cornagraphic hole was baffled, it reduced the amount of 
light from the primary by a factor of 6 at 1$\arcsec$ compared to direct 
imaging (Schnieider et al 1998). 
Finally, there are two main filters on Camera 2 which could have been used, 
F165M (1.55--1.75\micron) and F160W (1.4--1.8\micron). There was a
question which of these would be more useful in detecting very cool
objects (T$<$1000K) because of the presence of methane absorption at
H-band. Using the 
spectrum of Gl 229B, a comparison of the filters found that, a similar
flux and temperature object would have 1.6 times greater signal 
to noise ratio in the wider filter. Therefore, we conducted 
our imaging survey on Camera 2 with the coronagraph at F160W.

In target acquisition the centroid of the target's PSF 
is placed at the 'low-scatter-point' of the hole by an on-board
acquisition. It is slightly different from the center ($-$0.75 pixel in
X and $-$0.25 pixel in Y) to optimize stray
light rejection determined by in-flight testing of the 
NICMOS IDT. (Schneinder et al. 2001). Our choise of the centration points within the 
coronagraphic hole was designed to reduce the intensity of the speckle pattern 
from the optics at the edge of the hole. 

With the target behind the coronagraph, we reduce the amount of
light in an azimuthal average at the edge of the coronagraph 
compared to direct imaging allowing for deeper investigation for 
comapnions. We developed an observing strategy to go even deeper, as
explained below.

\subsection{Roll Subtraction}

The Hubble telescope has the ability to slew in three ways:
in x, in y, and in $\theta$. In order to keep the target star in the same
position behind the coronagraph, as well as simplify the removal of the
stellar profile, we designed two observations of the star at 
different spacecraft orientations ($\delta\theta$) separated by 29.9
degrees\footnote{A few observations were made at $\delta\theta <$29.9
due to lack of sufficient guide stars at both orientations}. This was
the maximum roll the telescope could maintain and still keep its solar
arrays pointed toward the sun. For a companion
0$\farcs$5 from the star, a 26 degree roll is the minimum needed for
the two PSF's (the companion's PSF at the two roll angles) to be separated 
by a distance equal to their first Airy maxima. 
Therefore, the observing strategy is to place the star
behind the coronagraph, observe for about 800s, roll the telescope, 
and observe again for 800s. This integration time kept both rolls 
to fit into one orbit, allowing for the roll and second target acquisition. 
Actual integration times per observation 
varied from 42s to 256s depending on the brightness of the primary 
(saturation at 0$\farcs$4 was avoided) 
with multiple frames adding up to $\sim$800s). The 800s was 
typically split into at least 
three Multi-Accum (non-destructive read) integrations (MacKenty et al. 1997). 

\begin{figure}
\figurenum{2}
\includegraphics[scale=.60,angle=270]{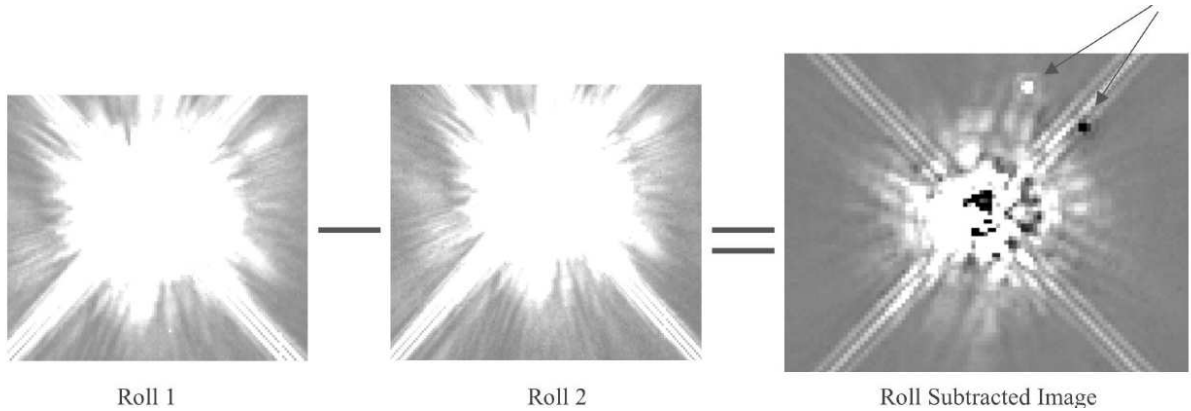} 
\figcaption{The roll subtraction is very effective in eliminating much of
the light from the primary star (ex: TWA 7). The star was imaged in the
coronagraphic hole (a), the telescope was rolled by 29.9 degrees, and the
star was imaged again (b). Subtraction (c) of the two images reveals a
stellar-like object approximately 2$\arcsec$ and 10 mags fainter than the
primary. (All three images are stretched from $-$1 to 1 adu s$^{-1}$ and are
approximately the 4$\arcsec\times$4$\arcsec$ around the coronagraph). The object 
around TWA 7 was found by follow-up observations to not have the same 
proper motion as the primary.}\label{rlsub}
\end{figure}

This helped reduce intrinsic detector 
non-linearities and lessened the chance of saturating pixels 
at small angular distances from the
occulted core of the target PSF. It also made maximum use of the available
dynamic range. When we subtract the two images,
the background from the star should ideally subtract out 
and if there are any objects are in the field, two images of
each should remain (Figure \ref{rlsub}). 
Because of thermal induced changes in the telescope 
(or 'breathing') the HST+NICMOS PSF varies on close to an orbit timescale 
(even slightly within an orbit) and so scattered light residuals persist 
as the 10\% of the light not contained in the coronagraph will change slightly 
(Schneider et al. 2001).
Therefore, observing at two rolls within an orbit mimimized these changes. 

\subsection{Reduction of NICMOS data}

The NICMOS coronagraphic images were independently reduced and processed 
utilizing calibration darks and flat-fields created by the 
NICMOS IDT (Instrument Definition Team) from on-orbit observations, rather than 
library reference files prepared by STScI. Mainly, flatfields were augmented 
with renormalized data from contemporaneous lamp calibration images (obtained as 
part of the acquistion process) to help remove artificial edge gradients and 
allow for calibration directly around the coronagraphic hole (Schneider et al. 2002).   
Using the NICRED program (McLeod 1997), the observations were 
bias and dark subtracted, corrected
for non-linearities using the ramp of the multiaccum imaging mode, and flat-fielded. 
Bad pixels were 
masked in the calibration steps and replaced with values interpolated from neighboring 
pixels with a cubic spline function. 
The three calibrated images at each orientation were then averaged to 
create final images.

As a final step, the cleaned calibrated images from each of the 
two spacecraft orientations 
were subtracted from each other leaving lower amplitude residual noise near the
coronagraphic hole edge, as well as positive and negative conjugates of any 
objects in the field of view (Figure \ref{rlsub}). Registration of the 
images was done 'by-eye' for the smallest residuals 
as the second image's positon was changed 
by 0.05 pixel in the X and Y directions. It was found 
early on that the diffraction spikes were never quite aligned in 
the best azimuthal subtraction, and 'by-eye' ultimately 
led to the best subtraction. A significant component of the 
residuals in the diffraction spikes arise in a small misalignment 
of the image of the mirror support structure formed in the pupil plane and 
small shifts of the pupil mask itself over time, inducing a diffraction 
of the spikes themselves (Schneider et al. 2001). The difference image, therefore, produces a 
'triple' diffraction spike that varies in amplitude and phase, but 
has nothing to do with centration. 

Since the target star is occulted in the coronagraphic images, its position to 
measure offsets of possible companions must be ascertained indirectly.  
First, the target's position was found in the acquisition image by a least-squares isophotal
ellipse fitting process around the PSF core with a radius of 7 pixels to
exclude flux from any close objects.  The target placement in the
coronagraphic frames was then determined by applying the target-slew
vectors used by NICMOS, accessible from the HST engineering
telemetry, resulting in independent measurements of each offset and 
position from both orientations.

\section{Point-Source Detection}
\subsection{Determining Detection Limits}

For a full analysis of the results, 
it is necessary to determine sensitivity to point sources. 
To determine the limits at which sources could be detected
within the NICMOS roll-subtracted images of the observed stars, we 
planted\footnote{software courtesy of A. Ghez and A. Weinberger.} 
point-spread-function (PSF) stars, 
generated with Tiny Tim (Krist \& Hook 1997), 
at random locations in every image. 
These PSF stars are noiseless, and can be
adjusted in flux. We examined a range of magnitudes from H=10--22,
stepped by 0.2 mag, for each roll subtraction. For each magnitude, 
25 stars were planted randomly within 3$\arcsec$ of the center of
the primary. This number of stars was 
chosen to avoid confusion between planted PSF's. For better statistics, it was
repeated 40 times for a total of 1000 planted PSF's at each
magnitude. Each of the 40 images was cross-correlated with the Tiny Tim
PSF to locate planted sources.
The values in the correlation map range from 1 for
perfectly correlated points to -1 for perfectly anticorrelated points.
The results are then compared with the log of the actual planting,
and the correlation coefficients of the planted stars are recorded, binned, and
averaged as a function of radius. 
Correlation coefficients of 0.9 or above are treated as definite detections.
This limit was very conservative set by the level above 
which no false hits (glints, etc) were found 
for several test images. As a test, two observations with candidate
companions in the field were also placed through the
routine; one with a H=17 ($\Delta$m=9.5) 
object at 2.5$\arcsec$ and one with H=20 ($\Delta$m=13) object at 2.5$\arcsec$. 
The brighter object was found with a correlation of 0.99, while the
fainter was found with correlation of 0.93. The
routine found nothing else in each image with a correlation
greater than 0.9. The separation from the star at which the correlation 
reaches 0.9 is then recorded for each magnitude step.

The F160W filter is $\sim 30\%$ wider than ground-based Johnson H-band
filters which necessitates a careful conversion from F160W to H band for
cool temperature objects. For six M 
dwarfs between spectral types M6 and M9 with measured
F160W and ground-based H-band magnitudes, we find a mean difference
of 0.03 mag. 
Objects with a spectrum like Gl 229B, displaying methane absorption, have a
H$-$[F160W] color of about 0.2 mag. The quoted H magnitudes of objects
less than 80 M$_{Jupiter} $\ are uncertain by 0.25 mag due to uncertainties in
the opacities and gravity dependence of methane absorption 
(A. Burrows, pers comm).

\begin{figure}
\figurenum{3}
\epsscale{0.8}
\plotone{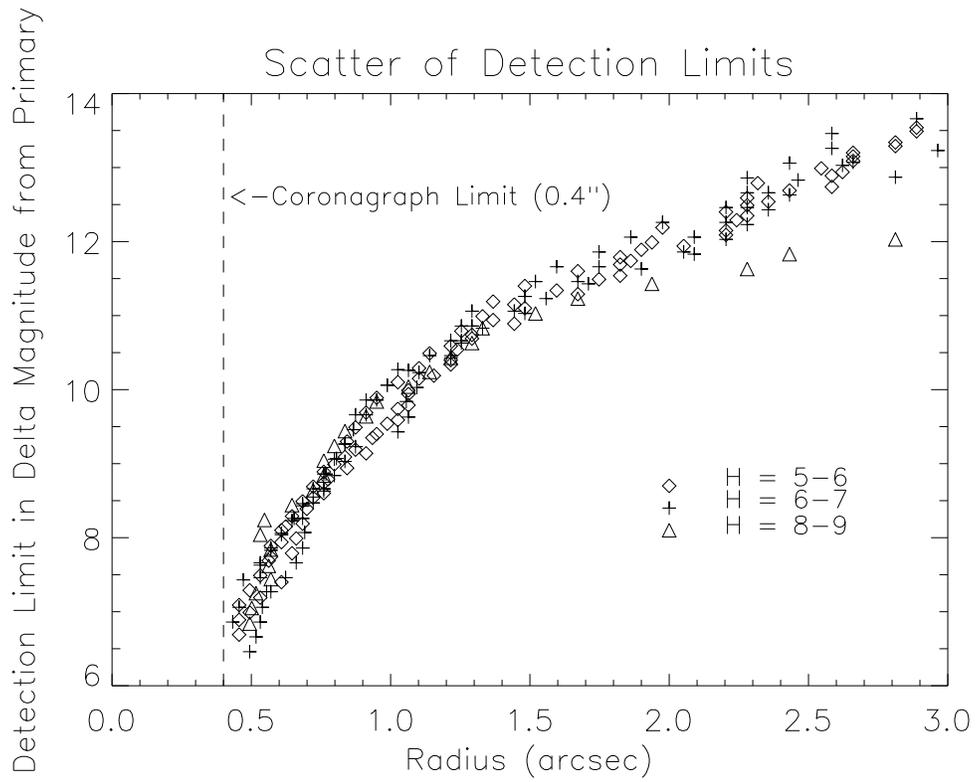}
\figcaption{Detection limits found by planting and recovering PSF stars
in a range of magnitudes within 3$\arcsec$ of the center of
the coronagraph in the roll-subtracted images. Different symbols 
indicate the different brightness of the primary star plotted.}\label{scatter}
\end{figure}

In figure \ref{scatter}, we plot the detection limits found overall in
the observations. 
Our sample has an average primary magnitude H $=$ 7 mag and a median age of 0.15
Gyr. At 1$\arcsec$, we can confidentally detect a delta magnitude of 9.5 mag for all 
stars. At a median
distance of 30 pc, 1$\arcsec$=30AU and our average limit corresponds
to M$_H$ = 14.1 mag. From the
models of Burrows (pers comm), this corresponds to less than 20 M$_{Jupiter} $. 
The most distant stars are the TWA association, in which 1$\arcsec$
correlates to 50 AU, but they are much younger (10 Myr), so the
detection limit is down at a few M$_{Jupiter} $. At 0$\farcs$5, we can detect a
delta magnitude of 7 mag, approximately 3 mag better than most speckle
imaging programs (i.e. Bouvier, Rigaut, \& Nadeau, 1997; Patience et al 1998). 

This program fully sampled 45 young stars with a median age of
0.15 Gyr with the ability to detect 30M$_{Jupiter} $\ brown dwarfs. 
For several of the younger stars
we were able to probe deeper in the mass range to only a few M$_{Jupiter} $. For
the median distance of 30 parsecs, our search covered 12-120 AU in
orbital separation. This covers the maximum of the distribution of
companion separations found in the open clusters Pleiades, Alpha Persei and
Praesepe, the G-dwarf radial velocity study and observations for T
Tauri stars (Patience, J. 1999). 

\subsection{Detectability of high-mass planets}

\begin{figure}
\figurenum{4}
\epsscale{0.8}
\plotone{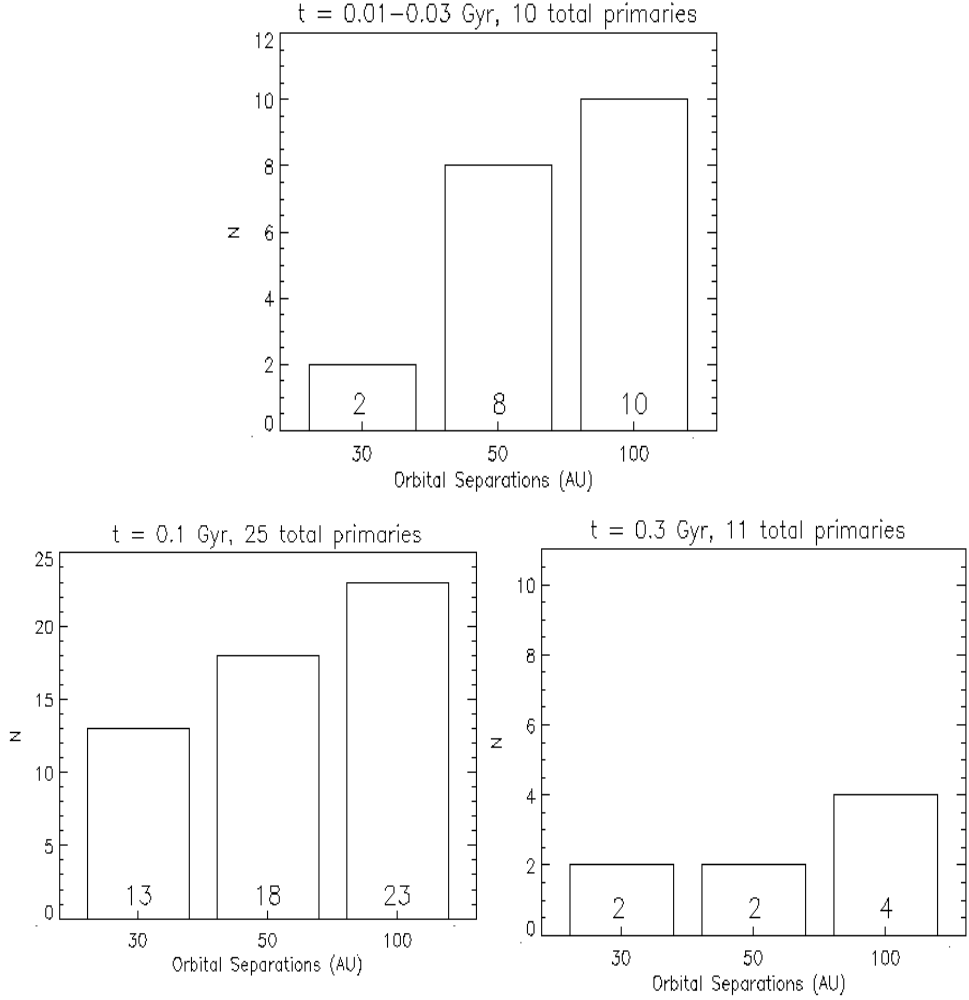}
\figcaption{The number of primaries, grouped by age, 
in which a 5 M$_{Jupiter} $\ object could have been detected at the 
orbital separations of 30, 50, and 100 AU. Based on the models of
Burrows (pers comm), a 5 M$_{Jupiter} $\ object was assumed to have an absolute 
H magnitude of 14.2, 16.7, and 18.7 mag, at the ages of 
0.02, 0.1, and 0.3 Gyr, respectively. This study explored the 
high mass planet range above 5 M$_{Jupiter} $\ around 36\%, 61\%, and 80\% of the 46 
primaries at 30, 50, and 100 AU, respectively.}\label{planets}
\end{figure}

At large 
separations from the primary, we were able to detect objects into the high
mass planet range. For our oldest stars, t=0.3 Gyr, a 5 M$_{Jupiter} $\ object is
expected to have a absolute H mag of 18.7 mag (Burrows, A. pers
comm). Outside of 5$\arcsec$, for the majority of stars, detection was 
limited by the sensitivity of the images, [F160W] $\sim$ 22, set by
the integration time, so a 5M$_{Jupiter} $\ object was detectable for primaries at
30 parsecs. Closer 
than 5$\arcsec$ detectability depends on the brightness, age, and distance 
of the primary. Using absolute H magnitudes of a 5 M$_{Jupiter} $\ object of 
14.2, 16.7, and 18.7 mag, at the ages of 
0.02, 0.1, and 0.3 Gyr, respectively, we plot the number of
primaries in which a detection was possible at separations of 30, 50,
and 100 AU (Figure \ref{planets}). This survey could have detected a 
high mass planet above 5 M$_{Jupiter} $\ around 36\%, 61\%, and 80\% of the 46 
primaries at 30, 50, and 100 AU, respectively. 

Radial velocity surveys have found $\sim$ 6\% of the over 1000 
FGKM stars have high mass planets within a few AU (Marcy et al. 2004). 
Since Jovian planets are
thought to form at larger distances, several theories have been
proposed to explain how these massive planets are so close to their
stars. One mechanism proposed (Lin \& Ida 1997) is that multiple planets
forming within a disk interact gravitationally to kick one planet to
an eccentric orbit inside 1 AU, seen in radial velocity studies, 
while causing another to be ejected to
much larger radii ($>$50 AU). 

It is not a strong constraint, but for the 45 stars 
this program observed, detecting 1--2 high mass planets at large radii
would be consistent with all of the radial velocity planets having formed in
the interaction method described above. However, it is not possible
with the imaging data at a single epoch to distinguish between objects
formed in situ or those kicked out.

\section{Identification and Analysis of Point Sources}

In Table 2 we present the results of point sources identified in the coronagraphic
survey. `Stellar-like' candidates are those that have a FWHM between
0$\farcs$14 and 0$\farcs$18, and usually show an Airy diffraction pattern. The
high resolution of the observations 
makes it easy to distinguish between stars and 
diffuse background galaxies. Several of the candidate companions 
have been re-observed for proper motion and spectra 
to confirm companionship and their possible substellar
nature. Many candidates were very close to their primaries and  
had to be followed up with other space-based observations or adaptive optics. 
In the next sections we will 
describe the further observations on the candidates that confirmed their background nature 
or companionship.

\subsection{Observations}

\subsubsubsection{STIS}

The candidate secondaries around HD 177996, TWA 5, HR 7329, GL 577, HD 102982, 
and GL 503.2  were observed between March and July 2000 
with STIS (Program 8176).
Each primary was acquired into the 52$\arcsec\times$0$\farcs$2 slit 
and then offset based on the NICMOS
astrometric results to place the secondary into the slit. To keep the
primary as far out of the slit as possible, we employed a slit
position angle 
so that the line joining the primary and secondary was approximately 
perpendicular to the slit, thereby minimizing contamination from 
scattered primary light. Spectral imaging sequences were
completed in one orbit per star with the G750M grating in three tilt settings
with central wavelengths of 8311,
8825 and 9336\AA\ (resolution $\sim$ 0.55\AA).For HR 7329B, HD177996, Gl503.2B and 
HD 102982B we integrated for 684s, 344s, and 310s respectively at the three 
tilt settings and for GL 577B and TWA 5B the integration times were 340s, 
172s, and 174s. For TWA 5B and Gl 577B, we obtained flat fields after each
set of four spectral images. These were recommended by the STIS team
to calibrate the known effects of fringing which appear longward of
$\sim$ 7500\AA\. After our first sets of medium resolution spectra
were reduced, the library flats served as well as these
contemporaneous flats shortward of $\sim$9000A. For the rest of the
observing program, we used library flats
and no longer took flat fields on-orbit. This allowed more time for
integration in the later spectra.  
At each tilt setting we executed a two position dither of 0$\farcs$35 along the
slit to allow replacement of bad or hot pixels, and the 
exposures were split for cosmic ray removal. Thus, we obtained four spectra
at each tilt setting.

\subsubsubsection{CFHT}

The Gl 577, Gl 503.2 systems were observed on 4 and 5 Mar 1999 UT, 
and HD 180445 was observed 11 Nov 2000 with the Canada-France-Hawaii 
Telescope (CFHT) using the adaptive optics (AO) system PUEO 
(Rigaut et al. 1998) and its 1024x1024 HgCdTe Hawaii detector KIR 
(Hodapp et al. 1994). PUEO delivers diffraction limited 
images at near-infrared wavelengths (PSF FWHM of 95, 110 and 140 mas for J, 
H and K bands respectively). KIR is used at the 
F/20 output focus of PUEO and is sensitive in the [0.7-2.5]$\micron$
range. Despite poor meteorological conditions, with cirrus covering most of the 
sky and an uncorrected seeing varying between 0$\farcs$9 
and 1$\farcs$5 over the nights, PUEO 
was able to easily detect the bright companions of HD 180445, Gl 577, and Gl 503.2 
in a single exposure of a few seconds. All three systems were imaged in the J and K
bands.

After correction for bad pixels and flat-fielding procedures, we subtracted the
sky background. For this purpose, a series of images was obtained for each filter 
by placing the science target near the center of each of the detector quadrants. 
The images were co-added, and the sky background was derived from the median of 
this set of four images. 

\subsubsubsection{Palomar}

The stars Gl 577 and Gl 503.2 were observed on 14 May 2000 UT, and 
the stars RE0723+20, GL 875.1, and GL207.1 were observed on 
26 September 2000 UT with the 200 inch Hale telescope at 
Palomar Observatory using the AO system PALAO and its
1024x1024 HgCdTe Hawaii detector, PHARO (Hayward et al. 2000). 
The system mounts at the Cassegrain focus and can achieve resolution of
0$\farcs$05 at K band. 
Both Gl 577 and Gl 503.2 systems were observed at K
band, and Gl 577 was additionally observed at H. 

Basic reduction included correction for bad pixels and flat-fielding procedures 
following  the same methods for the CFHT observations. 
Clouds precluded direct photometric observation against standards. 
We measured the separations for each observation of the pairs. Using 
the neutral density filters for the position of the saturating primary 
when the secondary is not seen may lead to a larger statistical offset. 

\subsubsubsection{Keck}
The Gl 577 system was observed on 12 Aug 2000 using the AO 
system  with the slit-viewing camera (SCAM) on NIRSPEC (McLean
et al. 1998). The H-band Strehl
ratio, as measured on the primary, was 0.16 with
FWHM$=$0$\farcs$045 in a 10 second exposure. 

Due to the small field-of-view of the infrared camera (5$\arcsec \times
5\arcsec$), the primary and secondary were observed in different frames, but the
adaptive optics atmospheric corrections were made on the 
primary during the entire observation.
Six H-band images were obtained of the secondary pair in sets of 10
coadds of 1 second integrations dithered about the array.

\subsection{Analysis of Follow-up Observations}

\subsubsection{Background Objects}

Indeed, when searching for companions, not all point sources will be companions. That was 
true with several of the objects detected in this survey. They are presented here as a 
time-saving service for future astronomers. 

\subsubsubsection{Candidate companion to LP 390-16 is a background object}

Analysis of the roll-subtracted, coronagraphic images of LP 390-16 reveal a 
stellar-like object at a separation of 1.45$\arcsec$ $\pm$ 0.08, and a position angle of 
226$^{\circ}$ $\pm$1 with a F160W magnitude of 14.4 $\pm$ 0.1 mag.  

To determine possible companionship with the primary, the 1954 epoch Palomar 
Observatory Sky Survey (POSS) digitized plate was examined in which a star (star X) 
was found located approximately 10$\arcsec \pm$1 at a 
position angle of 100$\pm$5 degrees from LP 390-16, or 10$\arcsec$ east and 
2$\arcsec$ south. With a 44 year baseline, if the two stars 
are unassociated, then LP390-16, with a proper motion listed in table \ref{pmA},  
should move approximately 9$\arcsec$ 
east and 1$\farcs$5 south leaving a separation of approximately 1$\farcs$1
at a position angle of 120 degrees if star X is background. This is close to the 
detected point source within the errors of measurement on the POSS plates, 
and the candidate companion was the only object near this position 
in the NICMOS images. Therefore we conclude the secondary we detect in 
the NICMOS images is consistent with this same background star. This object is 
not seen in the second epoch POSS plates, epoch 1993, which could be due to the 
current proximity of LP 390-16.

\subsubsubsection{Candidate companion to HD 177996 is a background object}

A STIS follow-up observation of the candidate companion of HD 177996 was 
attempted on 27 Feb 2000. In the NICMOS image the m$_H$=19.1 candidate was 
located at 5.17$\arcsec$ at a PA of 300$\fdg$1. A spectra of the primary star 
was obtained, and then at the astrometric offset, a spectra of the 
secondary was attempted by HST. No detection of the secondary was made in a 
total exposure time of 10906 seconds with the G750L filter, 
though the diffraction spikes of the primary are visible at the 
appropriate separation indicating the offset was performed 
correctly. HD177996A has a fairly decent proper motion (Table \ref{pmA}) 
and the NICMOS detection was almost two years earlier (3 May 1998).  
The half-width of the STIS slit is 0$\farcs$1, so if the candidate companion 
is a background star it could easily have been at the edge or out of the slit. 
We therefore conclude the object is a background star and 
not a companion to HD 177996.

\subsubsubsection{Candidate companion to RE0723+20 is a background object}

After the NICMOS observation on 20 October 1998, RE0723+20 was reobserved 
on 26 September 2000 with the Palomar AO and the closest of the point sources was 
seen and measured to be separated by 5.4$\arcsec$ $\pm$0.1, 
and a position angle of 326$\fdg$4 $\pm$0.4. With a proper motion listed in 
Table \ref{pmA} an object at the original separation and 
position angle (Table 2) is expected to be at a separation of 
5.47$\arcsec$ and 326$\degr$ in 
twenty-three months if it is merely a background object. 
We conclude this object, therefore, is not a 
companion. The other two point sources seen in the 
HST field of view (FOV) were outside the Palomar AO FOV.

\subsubsubsection{Candidate companion to LHS 2320 is a background object}

On 01 April 1998, one of the point sources was observed at a separation of 
14.92$\arcsec$ $\pm$0.11 and a position angle of 96.9$\degr$ $\pm$ 0.2 from 
the M5 star LHS 2320. Searching the 2MASS  point-source database we find  
two similar sources at the position of LHS 2320 that 
were observed on 12 Feb 2000 and separated 
by 16.2$\arcsec$. With the change in separation over the 1.8 year baseline 
similar to the primary's proper motion (Table \ref{pmA}), no other point sources in 
the vicinity of similar brightness, and a J-K = 0.38 mag color, 
we conclude it is most likely a background 
G star (10:52:15.3 +05:55:08). 

\subsubsubsection{Candidate companion to Gl 875.1 is a background object}

After the NICMOS observation on 7 Jul 1998, Gl 875.1 was reobserved 
on 26 September 2000 with the Palomar AO and the point source was 
measured to be separated by 7.1$\arcsec$ $\pm$0.1. With the star's 
proper motion (Table \ref{pmA}),  a background object at the 
original separation and position angle is expected to be at 
a separation of 6.5$\arcsec$ and 255$^{\circ}$ within a time baseline of 
twenty-three months. We conclude this object, therefore, is not a 
companion.

\subsubsubsection{Candidate companion to TWA 7 is a background object}

TWA 7 was observed with the NICMOS coronagraph on 1998 March 26 
with a candidate companion at 
a separation of 2.44$\arcsec$ $\pm$ 0.05, and a position angle of 
142.2$^{\circ}$ $\pm$ 0.1. TWA 7 was also reobserved 
on 02 November 1998 (Program 7226) with the NIC 1 (pixel scale  = $\sim$0.043
arcsec pixel$^{-1}$) camera with a
medium-band F090M filter (central wavelength: 0.9003~$\mu$m,
$\Delta\lambda$ = 0.1885~$\mu$m). It was observed in
a 2 position dither with 4 Multi-Accum integrations of 64s taken at each
position for a total of 512s. These were dark subtracted and
flat-fielded using calibrations created by the NICMOS IDT.

The F090M filter is a red I-band, covering from 0.8$-$1.0$\mu$m. The
colors of the possible companion (F090M-F160W=0.72) are consistent 
with a mid-K star. The candidate was observed with Keck on 20 February 2000 
and had changed separation from the primary from 
2.44$\arcsec \pm$0.05 to 2.54$\arcsec \pm$0.08 over the almost two year 
baseline. With the the change in separation and color, we conclude TWA 7 
and the candidate were not associated, in agreement with Neuhauser et al (2000a)

\subsubsubsection{Candidate companion to TWA 6 is a background object}

A point-source (H=19.93 $\pm$ 0.08 mag) was discovered on 20 May 1998 at 
a separation of 2.549$\arcsec$ $\pm$ 0.011, and a position angle of 
278.7 from the young star TWA 6. The field was re-observed with NICMOS 
on 2002 June 10. The point source  
was found to lie at a separation of 2.356$\arcsec$ $\pm$ 0.009 and a position angle 
of 281.5$\degr$ $\pm$ 0.1 from TWA 6. This change over the 4 year baseline 
corresponds to a $\delta$ (RA,Dec) = (-52.2, -21.2) mas/yr. Webb et al (1999) reported the 
proper motion of TWA 6 to be (-60,-20) mas/yr, which makes this change in 
separation wholly consistent with being a background object and not a companion 
object. 

The field of TWA 6 was reobserved with the NICMOS 1 camera on 09 December 1998 
(Program 7226) with a
medium-band F090M filter and the point source was not detected. 
We derived an upper limit (3$\sigma$) to the flux of [F090M]=22.6 mag 
in the predicted position from the NICMOS images. 
Using low-temperature models to transform between F090M and I-band, we 
calculated an upper limit of I$-$H$>$3.3 for the candidate companion. 
We conclude the object is a very red background object, nonetheless.

\subsubsection{Confirmed Companions}

In this section, we discuss several positive confirmations from this
survey. Two of the discoveries, TWA 5B and HR 7329B have already been presented in 
L99 and L00 respectively. Independent proper motion confirmations for 
TWA 5B, HR 7329B and GL 577B have been presented in Neuhauser et al (2000b), 
Guenther et al. (2001) and Mugrauer et al. (2004). 
We present the results of the follow-up observations for TWA 5B and 
bright point sources around other stars with fairly well-known ages.

\subsubsubsection{Bright companions near HD180445, HD 220140 and HD160934}

Analysis of the subtracted, target-acquisition images of HD 180445 observed on 27 August 1998 
reveal a stellar-like object at a separation of 9.45$\arcsec$ $\pm$ 0.08, and a position angle of 
52.5$^{\circ}$ $\pm$0.5.
On 11 November 2000, the point source was re-observed at 9.5$\arcsec$ $\pm$ 0.1 and a position angle of 
53.0$^{\circ}$ $\pm$0.8 in the 0.5s acquistion image before acquiring HD 180445 behind the 
coronagraph on CFHT. The proper motion of HD 180445 (Table \ref{pmA}) should have 
changed the separation in two years by 0.27$\arcsec$ if it was a background object. 
From the 2MASS database we find 
a point-source at a separation of 9.47$\arcsec$ $\pm$ 0.02, and a position angle of 
53.0$\degr$ $\pm$0.1. From both of the observations and the 2MASS observation, 
we conclude that this pair is most likely a common proper motion pair within the errors. 
The 2MASS magnitudes of HD180445A are J=6.99$\pm$0.02 mag, H=6.49$\pm$0.03 mag, 
and K=6.38$\pm$0.02 mag, while the magnitudes of HD180445B are J=10.46$\pm$0.04 mag, 
H=9.88$\pm$0.04 mag, and K=9.69$\pm$0.03 mag. The J-K color of 0.77 and derived absolute 
magnitudes are consistent with an early to mid-M star.

A point source was found at a separation of 10.85$\arcsec$ $\pm$ 0.08, and a position angle of 
216.3$\degr$ $\pm$0.5 from HD 220140 when it was observed with NICMOS on the 25 October 1998. 
From the 2MASS database 8 Oct 2000, one point source was found within a 15$\arcsec$ search. The 
candidate companion is separated by 10.85$\arcsec$ $\pm$ 0.08 and a position angle of 
215.15$^{\circ}$ $\pm$0.05 from the primary. From the listed proper motion of HD220140 (Table \ref{pmA}) 
the two should have been separated by 11.20$\arcsec$ and position angle of 217.5$\degr$ 
during the two year baseline if the point source was a background object.  
Therefore the candidate is most likely a proper motion companion within the errors. 
The 2MASS magnitudes of HD220140A are J=5.90$\pm$0.02 mag, H=5.51$\pm$0.04 mag, 
and K=5.40$\pm$0.02 mag. The magnitudes of HD220140B are J=8.04$\pm$0.02 mag, 
H=7.39$\pm$0.03 mag, and K=7.20$\pm$0.02 mag with a J-K color and derived absolute 
magnitudes consistent of a mid-M star.

Weiss (1991) detected a second object at 20$\arcsec$ at a position angle of 150$\degr$ 
from HD 160934 which did not change position angle or separation from 1955 to 1990. 
From the 2MASS database observation on 12 May 1999, one point source is found at 19.06$\arcsec$ 
at a position angle of 151.1$\degr$ from the primary. In the 44 year baseline, 
with the proper motion of HD 160934 (Table \ref{pmA}), the separation would have increased by at 
least 1.5-2$\arcsec$ if it was not a companion. The 2MASS magnitudes of the 
companion are J=10.25$\pm$0.02 mag, 
H=9.67$\pm$0.02 mag, and K=9.42$\pm$0.02 mag with a J-K color (J-K=0.83 mag) and 
derived absolute magnitudes consistent of a mid-M star.

\subsubsubsection{STIS Spectra of Gl 503.2, Gl 577, TWA5B and HD 102982}

\begin{figure}
\figurenum{5}
\epsscale{0.7}
\plotone{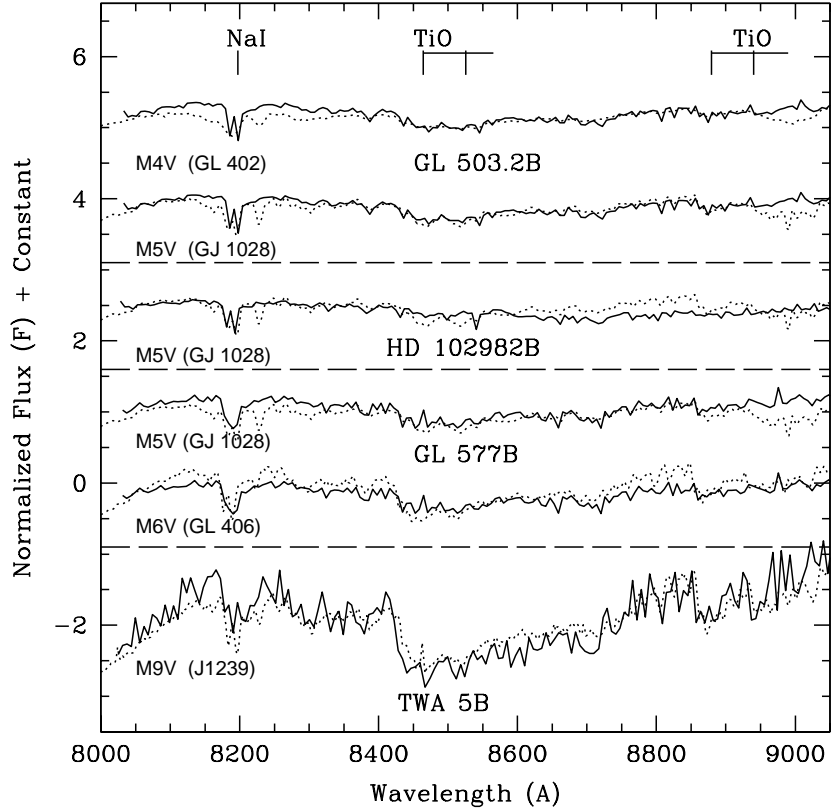}
\figcaption{STIS spectra of TWA 5B, Gl 577B, HD 102982B and Gl 503.2B (solid)
(normalized from ergs/s/cm$^2$/\AA) separated by the dashed horizontal lines,
compared with standard late-type M dwarf spectra (dashed)
(Kirkpatrick et al. 1991;Kirkpatrick et al. 1997). 
The zero level of each
spectrum is -3.3, -1.4, 0.2, 1.3, 2.7 and 4 respectively. The best fitted spectrum
was chosen with emphasis on the NaI absorption near 8200\AA\ and TiO
bands near 8450\AA\ . 
The best fit appears to lie between
M4V and M5V for Gl 503.2B, M5V for HD 102982B, between M5V and M6V for Gl 577B and 
M9V for TWA 5B. (The longward cutoff of 9050\AA\ is where the 
fringing affects the STIS spectrum.)}\label{gstis}
\end{figure}

Observations of four candidates were taken with the STIS instrument  
to determine if they were backgroud objects or companions. 
The STIS spectra were calibrated, averaged, binned to a resolution of
$\sim$ 6\AA\ and normalized to the flux (in ergs/s/cm$^2$/\AA) at 8500\AA. 
We compared the final, total spectra to those of standard 
low-temperature dwarf star spectra with a resolution = 18 \AA, 
a factor of three lower than our STIS spectra
(Kirkpatrick et al. 1991; Kirkpatrick et al. 1997) (see Figure \ref{gstis}). 
All spectra contain the Na I absorption doublet near
8200\AA\ which does not appear in
late-type giant stars. The TiO
absorption bands at 8450\AA\ and 8900\AA\ are also used to 
choose the best fitted standard. As seen in Figure \ref{gstis}, the slopes of the
spectra from 8600 to 8850\AA\ are small and fit the dwarf spectra well.
The best fit appears to lie between
M4V and M5V for Gl 503.2B, at M5V for HD 102982B,  
between M5V and M6V for Gl577B and M9V for TWA 5B.
(The spectrum for HR 7329B was previously 
published in L00). The spectral type derived 
for TWA 5B is consistent with photometric spectral 
type of M8.5 $\pm$ 0.5 spectral type derived by L99 
and Neuhauser et al (2000b).

\subsubsubsection{Binarity of the companion to Gl 577A}

\begin{figure}
\figurenum{6}
\epsscale{0.4}
\plotone{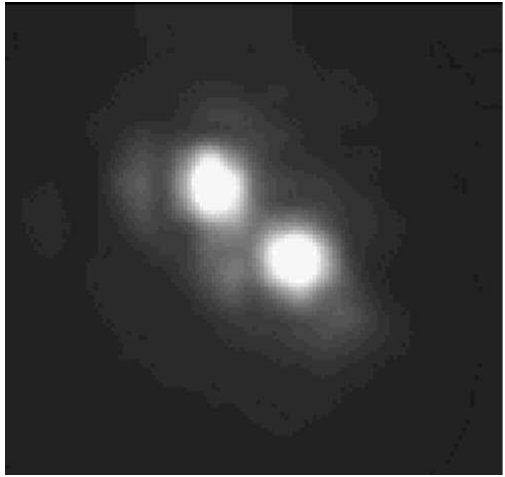} 
\figcaption{Infrared observations of the secondary Gl 577 B \& C taken 
with the AO system at Keck. The measured separation is 0.082$\arcsec$.
The orbit can be measured over several years to get dynamical masses for
the two components.}\label{577bin}
\end{figure}

The companion to Gl 577A 
appears to be elongated in both the CFHT images, Palomar, and the NICMOS images, 
suggesting that the companion is in fact a close binary system 
itself with an angular separation slightly lower than 
the instrumental resolution ($\sim$ 100 mas). The two components
were resolved with the Keck system (Figure \ref{577bin}). The
measured separation of 0.082$\arcsec \pm$0.005 corresponds to a
projected radial separation of $<$ 4 AU, 
corresponding to a period of approximately 20 years. With periodic
observations of these two components including radial velocity
measurements, the dynamical mass of this low-mass 
binary can be determined and used to check current evolutionary
models.

\subsubsubsection{Astrometry of the Companions}

\begin{figure}
\figurenum{7}
\epsscale{1.0}
\plottwo{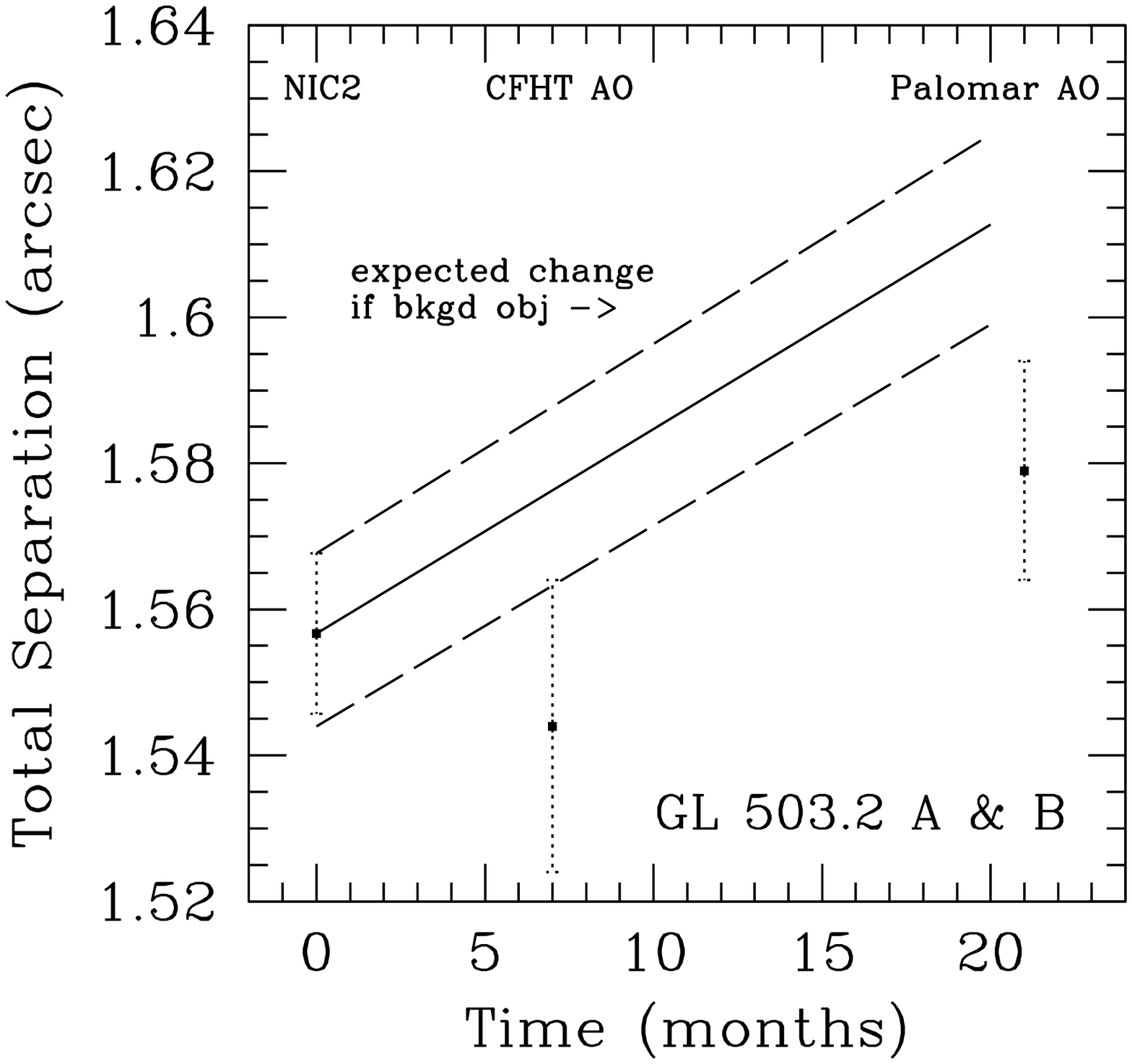}{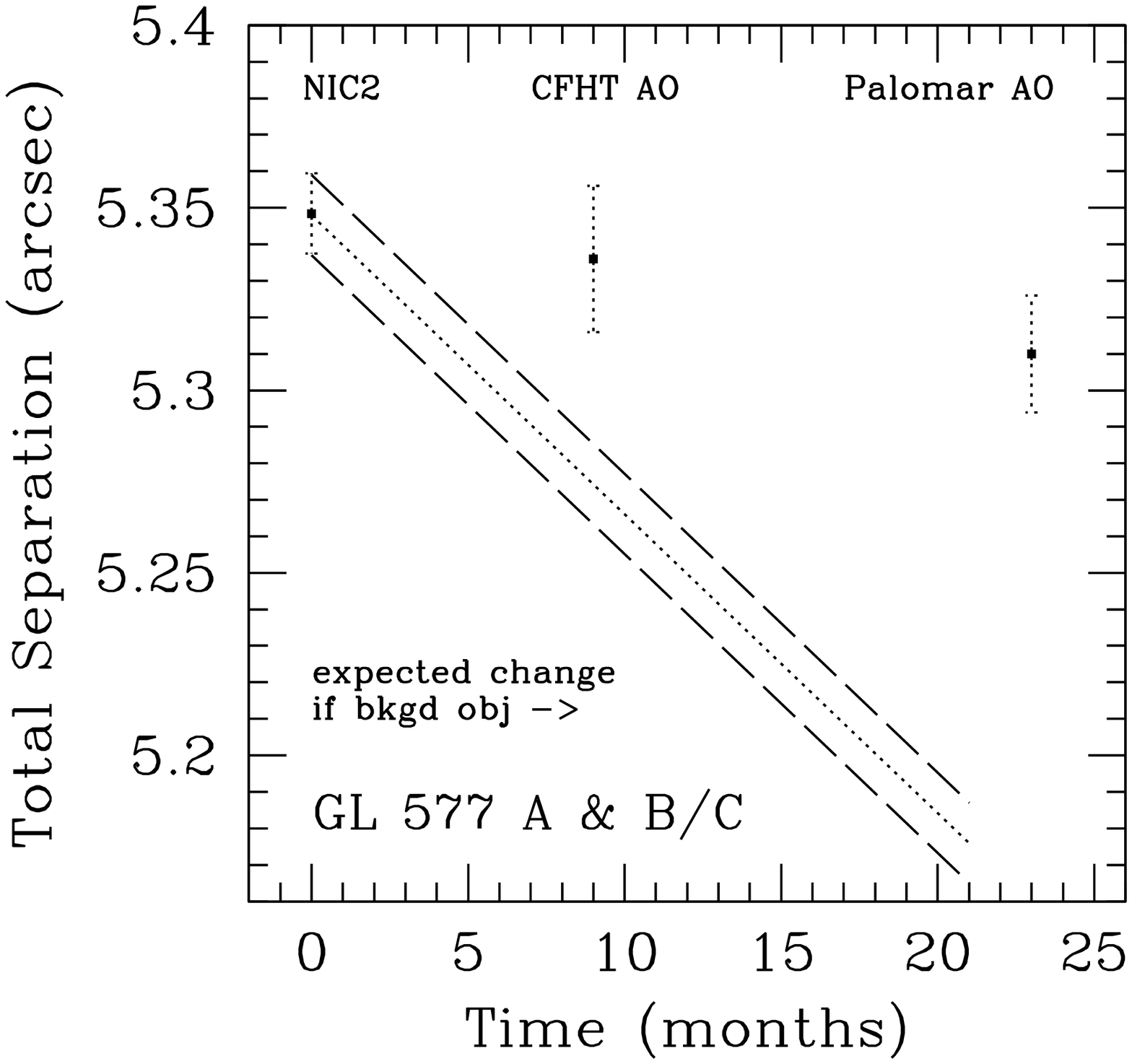}
\figcaption{Measurements of the separations of two companions 
(GL 577B/C and GL 503.2) made over
the last two years are compared to each other, taking into account a 1 sigma errors 
bar. The dotted line represents the change in separation
expected from the primary's known proper motion if it was 
not associated with the candidate companion.}\label{gstpm}
\end{figure}

From the ground based observations, a common proper
motion between the primaries and secondaries can be established.  
The known proper motions for the primary stars are listed in Table \ref{pmA}.
For Gl 577 and Gl 503.2 systems, the position of 
each object was measured in a gaussian centroid 
in each reduced image from NICMOS, CFHT and Palomar. 
The separations were then calculated and averaged over
all measures for each star at each epoch and plotted over time (Figure \ref{gstpm}).

If the secondary is not associated with the
primary, then the separation should change in an amount that is
calculable from the measured proper motion (Table \ref{pmA}). If Gl 503.2B is
not associated with Gl 503.2A, then the separation between them should
have changed from 1.557$\arcsec$ to 1.613$\arcsec$  from the NICMOS
observation until the Palomar observations. Instead, the separation
was measured at 1.579$\arcsec$. This is within the 1 sigma error of
the NICMOS measure, but 3 sigma outside of being a background
object. We therefore conclude Gl 503.2B is a likely companion of Gl 503.2A.

Similarly, if Gl 577 B \& C are not associated with Gl 577A, the
separation should have changed from 5.348$\arcsec$ to 5.176$\arcsec$
within the 2 years between observations. The separation was measured to be
5.31$\arcsec$ in the Palomar observations, three sigma above the 
expected change, and within 1 sigma of the NICMOS measure, leading to
the conclusion that GL 577 B \& C are indeed companions to Gl
577A.
 
The separation of HD 102982A and B was measured with NICMOS and 
again in the STIS acquisition before being placed in the slit. In the
STIS acquisition images, 
before and after centering, a 2d gaussian was fit to get a separation
of 0.943$\arcsec \pm 0.002$. 
A flux-weighted centroid using the DAOPHOT routine
received similar results. The position angle between the two
components was calculated to be 28$\pm$2 degrees. This is within
1$\sigma$ of the separation and position angle measured from the 
NICMOS image 2.25 years earlier. The proper motion
of the primary should change the separation by 0.21$\arcsec$, or
10$\sigma$, if the B
component was not associated, therefore we conclude HD102982B is a
companion to HD102982A.

\section{Determination of Mass}

Mass is the best determinant of the substellar nature of an object.
Unfortunately, the current ability to dynamically determine 
the mass of most these objects is impossible, therefore 
we must rely on evolutionary models of temperature, radius, and 
luminosity. To plot these objects on evolutionary 
tracks, we need the luminosity, or absolute
magnitude, and the effective temperature. We have converted NICMOS
measured photometry to absolute magnitudes using known distances 
from the primary, and used STIS spectra when available, and/or 2MASS infrared 
colors for the bright companions, to determine spectral type 
and therefore effective temperatures. Therefore, both parameters on an
H--R diagram have been determined independently, and have 
independent errors which are discussed below.

Proper motion has been demonstrated between the components Gl 577 
and HD 102982 within this paper. HR 7329 A \& B and TWA 5 A \& B have been
confirmed as a proper motion pair (Neuhasuer et al 2000; Guenther et al 2001).
Gl 503.2 A \& B are likely associated based on the density of background stars
and their measured properties. The luminosities or absolute magnitudes
of the secondaries have been derived using Hipparcos measurements of
the primary, which have determined distances known to within a few 
percent. This small error bar is taken into account when 
placing companions on the diagrams.
The parallactic distance was measured to the primaries by the Hipparcos mission
(Table \ref{cand}). With a derived H magnitude of 10.45 for Gl 503.2B, and 
a distance modulus of 2.05, we calculate an absolute H magnitude of
8.38 $\pm$ 0.09
mag, with an uncertainty that includes distance errors and NICMOS
calibration errors. In a similar manner, we derive an absolute H
magnitude of 8.58 $\pm$ 0.15 for HD 102982B and 7.83 $\pm$ 0.09 
for the Gl 577 B \& C pair. From the resolved Keck data, we
measure a delta mag of 0.1 $\pm$ 0.1 for the B and C components,
therefore, equal magnitude components will have absolute H magnitudes
of 8.58 $\pm$ 0.13 mag. The companions to HD 160934, HD220140 and HD180445 have 
absolute magnitudes of 7.72 $\pm$ 0.04 mag, 5.92 $\pm$ 0.04 mag and 
6.78 $\pm$ 0.04 mag, respectively.

\subsection{Effective Temperatures}

The effective temperatures have been determined from the spectra,
taken with STIS, from 8000--9000\AA\ which have several spectral
features including bands of TiO and Na which are specific to 
low-temperature spectral classes from M4V--M9V. The spectral class 
assignments based on these absorption features 
are good to 0.5 spectral type, which is the error we assign. Determining
the effective temperature adds the largest error because the
relationship between spectral class and effective temperature is
largely unknown for late-M type stars. Luhman \& Rieke (1998) derive a linear relation based on 
spectal class from Leggett et al's (1996) fit to synthetic models which agrees 
with the newer models produced by Leggett, Allard, \& Hauschildt (1998) 
within an uncertainty of about 100 K. This relation of
Luhman \& Rieke (1998) was used to determine the
temperatures of our candidates' spectral types and a 100K error bar covers a 
$\pm$ 0.5 spectral type. We calculate 3180K, 3010K, and 2840K for M4 V, 
M5 V and M6 V respectively.
For the brightest companions, HD160934B, HD180445B and 
HD 220140B, we assumed a spectral type of M4$\pm$1 from the 2MASS colors 
and derived absolute magnitudes and get a temperature of 3180K $\pm$ 200K.  

\begin{figure}
\figurenum{8}
\epsscale{0.8}
\plotone{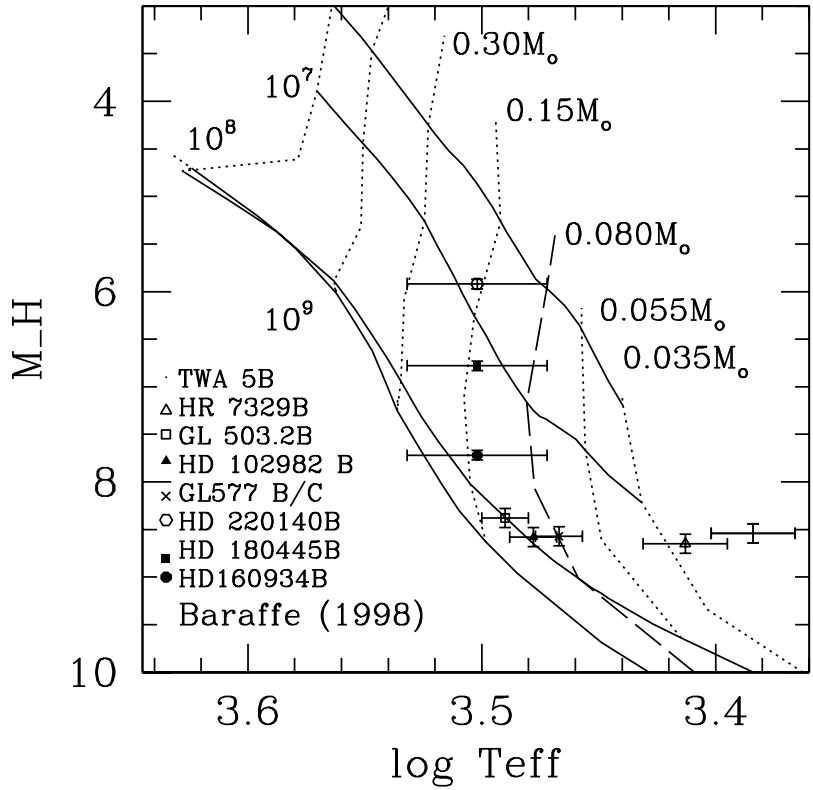}
\caption{H-R diagram of all eight of the discovered 
companions in this study using the absolute H magnitude determined distances 
of primary stars and effective temperatures derived from spectra or 2MASS colors. 
These models (Baraffe et al. 1998) indicate three low mass stars, 
two very low mass stars, a binary brown dwarf and 
two brown dwarf companions.}\label{gstevol}
\end{figure}

With the uncertainty for late M dwarf star
temperatures in mind, 
we plot the derived temperatures for each spectral 
class (Figure \ref{gstevol}) with their derived absolute H magnitudes.

\subsection{Derived Mass and Age}

From their placement on pre-main sequence evolutionary tracks (Baraffe et al 1998), 
we can infer a mass for the secondaries (Figure \ref{gstevol}). 
HD 220140B and HD160934B seem to be low mass stars falling near the 
10 Myr and 100Myr isochrone, consistent with derived ages (discussed in 
Appendix A) of their primaries. HD180445B appears to be a young low mass 
star near 50Myr, younger than the 
derived primary's age, which was an upper limit of 200Myr, especially 
since  Soderblom et al. (1998) speculated it might be a tidally 
locked binary and not young at all. 
HD 102982B and Gl 503.2B appear to be consistent
with very low-mass ($<$ 0.15 M$_{\sun}$) stars falling on the 100 Myr
isochrone with error bars extending between about 70 and 300 Myr. 
This agrees with ages derived for the primaries from other means. 
Gl 577 B \& C appears to lie just on 
the theoretical brown dwarf mass of 0.08 M$_{\sun}$ from its 
infrared magnitude and
spectral type at an age of about 70 Myr with error bars extending 30
and 100 Myr. This has been adjusted by 0.75 mag for being a binary. 
Since both components are of equal magnitude, this
would suggest both might be high mass brown dwarfs separated by $\sim$ 4 AU. 
The models would also suggest Gl 577 system is younger than the 
Pleiades (100 Myr) supported by the primary's chromospheric activity 
and rotation. Both HR 7329B and TWA 5B are bona fide brown dwarfs with masses 
of 30M$_{Jupiter}$ and 20M$_{Jupiter}$ as previously reported (L99;L00).

\section{Examining the Companion Mass Function}

Even though this survey was originally designed to discover brown
dwarfs, we can attempt to examine the companion mass function (CMF), and 
by extension the IMF, across the stellar/substellar border. 
Since masses can only be accurately determined for close binaries 
with measurable orbits, to derive appropriate masses
we use previously quoted models for the discovered companions.
In recent studies, in the field (Reid 1999), and the
Pleiades cluster (Zapatero Osorio et al. 1997), the relative number of low mass
stars and brown dwarfs per log mass interval is consistent with being equal,
suggesting a flat initial-mass-function ($\alpha=1$) for single stars. 

Several caveats exist in any statistical examination of a small size
sample. We have only examined a small separation range and thrown out
all known binaries within 20$\arcsec$, as mentioned in section 2. 
Earlier studies in
the Pleiades cluster and the field have used hundreds of stars while we have
observed just less than 50, which decreases the statistical significance. 
With these overall limitations in mind,  
we first estimate what one might expect 
in detections for our survey, by 
dividing the mass range into two logarithmically equal bins,
0.03--0.08M$_{\odot}$, and 0.08--0.2M$_{\odot}$. For an approximate idea of how many 
companions to expect, one can examine statistics within 5 parsecs of
the sun. Stars less than 0.2M$_{\odot}$ make up approximately 24 of 65 stars (37\%) within 
that distance range. Similar statistics are found in the nearest
100 stars taking into account missing systems. For statistics on the separation range of 30--120AU, 
one can look at the G-dwarf radial velocity survey and find that $\sim$30\%
of the stars have companions in that separation range (Duquennoy \& Mayor 1991). 

From these statistics, we might expect 30\% of our stars to have companions at the separations 
available within this study (30-120AU), and 37\% of those companions to be 
in the 0.08--0.3M$_{\odot}$ mass bin. Therefore we might expect 11\% of our 46 star sample, 
or 5 companions in this higher mass bin. This study found companions to HD 180445, HD 220140, 
HD 160934, Gl503.2, and HD 102982 that fit that expectation.

For the brown dwarf range (0.03--0.08M$_{\odot}$), we have three 
objects, HR 7329B and Gl 577B \& C. 
For an IMF with an equal mass of stars in equal bins ($\alpha=2$), 
we expect about 10 objects in the lower bin, which is
not consistent by 3$\sigma$ with our observations. 
For a flat IMF ($\alpha=1$), one which has equal number of stars in each mass bin, 
we predict 5 objects in the
lower mass bin. This is fairly consistent  with the findings of this
survey, though leans toward an $\alpha<1$. Therefore, even with the aforementioned caveats, 
the companion mass function we derive from this survey is 
consistent with the initial mass function found in the field (Reid
1999) and the Pleiades cluster (Zapatero Osorio et al. 1997). 

\section{Conclusions}

In the last decade, the search for  substellar objects  
has achieved spectacular success with the discovery of 
the first non-controversial detection 
of a brown dwarf, the definition 
of two new spectral types 'L' \& 'T', and the uncovering of over 
500 L \& T dwarfs. Even though observations of field dwarfs can 
answer many questions about formation and fundamental properties, 
many questions still exist that can only be answered by observations 
of brown dwarf {\it companions}. In order to directly detect 
substellar companions, we developed an 
infrared coronagraphic survey of young stars with the 
NICMOS camera on HST. 

Subtraction of two coronagraphic images taken within the same orbit 
at two angles of the spacecraft differing by 29.9$\degr$ 
was found to produce the most effective method of detecting point sources.
For the average primary magnitude of H=7mag, the survey detected a delta mag 
of 9.5 mag at 1$\arcsec$ with only the subtraction and no other manipulation 
of the images. This allowed detection into the high mass planet range at 
50 AU from over half of the primaries. Results from this survey 
include five low mass stars, 
two brown dwarfs, and one possible binary brown dwarf.

Models play an important role in substellar astronomy, and therefore 
every substellar companion discovered is extremely important.  
Because the primaries have been well studied, parameters such as age, 
distance and metallicities are known and 
can be used to provide fiducials to refine the present models. 
For example, the close binary GL 577 B \& C presented in this paper can be reobserved 
over the next few years to derive a dynamical mass to compare other young substellar 
field brown dwarfs. 
Higher resolution spectra in the visual and infrared of companions such as 
TWA 5B and HR 7329B are well constrained with age and metallicity and 
therefore provide anchors for models. Finally, as 
more substellar binaries and companions are discovered the parameters can be used 
as important pointers toward answering the questions of stellar formation versus 
substellar formation versus planetary formation.

\acknowledgments
This work was supported in part by NASA grants NAG 5-4688 to UCLA and 
NAG 5-3042 to the University of Arizona NICMOS Instrument Design Team.  
This Paper is based on observations obtained with the NASA/ESA 
Hubble Space Telescope 
at the Space Telescope Science Institute, which is operated by the
Association of Universities for Research in Astronomy, Inc. under NASA contract
NAS 5-26555. Some of the data presented herein were obtained at the W.M.Keck 
Observatory. We would like to thank the invaluable observatory staff at the 
CFHT, Keck and Palomar Observatories for their assistance. We would like to thank 
all the people who were helpful references during this study such as R. White, 
R. Webb, C. McCarthy, and R. Chary. We are grateful 
to the anonymous referee for a careful review of this manuscript and 
suggested improvements.

\newpage

\appendix
\section{Appendix A: Individual Target Ages}

\textbf{\textit{A-type stars}} It is difficult to determine 
ages for A-type stars, but HR 7329 \& HR 8799 appear to be
young ($<$ 40 Myr) based on rotation, and more importantly, 
location on an H$-$R diagram. 
For massive stars, rotational velocities decline with age; 
HR 7329 (A0V) has an especially large v$sini$ ($=$ 330 km/s) (Abt \& Morrel
1995) which is considerably above 
the majority of early A-type stars ($\sim$ 100 km/s). HR 8799 (A5V)
has a moderate v$sini$ ($=$ 40 km/s) (Abt \& Morrel
1995), and has been spectrally defined as a $\gamma$ Bootis star (Gray
\& Kaye 1999) which implies an age of a few to 100 Myr. 
A color-luminosity relation of nearby young clusters (L00) seems to
be correlated for stars of similar age; the 50-90 Myr IC2391 and
Alpha Per clusters lie below the older (600 Myr) Hyades and Praesepe. There is
a large scatter in the Pleiades (70-125 Myr), 
which could be due to a range of distances and 
ages as well as unresolved binaries.
HR 7329 and HR 8799 lie on a line located below the Alpha Per
and IC 2391 cluster which intersects $\beta$ Pic, HR 4796 and HD 141569, 
which suggests that HR 7329 and HR 8799 are between 10 and 30 Myr old (L00). 
Finally, it has recently been suggested that HR 7329 is a member of a young
co-moving cluster (Tucanae) much like the TW Hydrae Association with an age of $\sim$40 Myr 
(Zuckerman \& Webb 2000; Webb 2000).

\textbf{\textit{F-type stars}} HD 35850, HD 209253, \& SAO 170610 
are believed to be young ($<$0.3Gyr) due to their rotation activity, lithium abundance
and X-ray activity. 
Tagliaferri et al. (1994) use the Einstein satellite to measure 
X-ray luminosities of log(L$_x$)$=$30.0 erg s$^{-1}$ and 29.7 erg s$^{-1}$ 
for HD 35850 and HD 209253, respectively. Late F and G type stars in the Pleiades
typically have X-ray luminosities of log(L$_x$)$=$30 erg s$^{-1}$, 
while Hyades members fall under log(L$_x$)$=$29 erg s$^{-1}$. 
Tagliaferri et al. (1994) 
find for HD 35850 and HD 209253 vsin$i$=50 \& 16
km s$^{-1}$ and log N(Li) = 3.2, 2.9, respectively,
by fitting observations with synthetic model
spectra. These rotation
velocities fall between those of 
Pleiades members and the UMa Group.  
The positions of these two late F-type stars (log T$_{eff}\sim$3.8) on a Li vs
T$_{eff}$ plot is consistent with an age close to that of the
Pleiades, but lithium ages for F-type stars are questionable and must be 
supported with other observations (Favata et al. 1993). 
The coronal activities (X-rays) 
and high rotational velocities are consistent with the 
lithium age of 100-200 Myr.  
SAO 170610 has a log N(Li)=3.75, which indicates a younger
age, between Tau-Aur and Pleiades, or 20--125 Myr. There is also a small
measured X-ray flux, (log fx/fv$=$-4.15; Stocke et al. 1991) consistent
with a young age.

\textbf{\textit{G-type stars}} Ten G-type stars were selected based on
their chromospheric activity, lithium
absorption, and rotation. The flux ratio of Ca II H \& K emission to the
bolometric flux, or R'$_{HK}$, and Li abundance, expressed log N(Li), 
is compared among stars of young clusters. 
Henry et al. (1996) selected several stars as ``very active'' from Ca II
emission including HD 202917 (logR'$_{HK}=-$4.06) and HD 180445
(logR'$_{HK}=-$3.90). The Ca II H \& K ratios imply ages for these
objects, if these are single stars, similar to members of the Pleiades
cluster, using the chromospheric emission-age relation of Donahue (1993):
\begin{equation}
log (t) = 10.725 - 1.334 R_5 + 0.4085 R_5^2 - 0.0522 R_5^3
\end{equation}
where t is age in Gyr, and R$_5$ is defined as R'$_{HK}\times10^5$. 
Soderblom, King, \& Henry (1998b) measure a Li
abundance in HD 202917 (log N(Li)=3.28) consistent with 
Pleiades stars of similar spectral type, but 
measure only an upper limit (log
N(Li)$<$1.61) for HD 180445. Neither
star rotates rapidly for G-type stars 
with vsin$i$=12 and 8 kms$^{-1}$ for HD 202917
and HD 180445, respectively. Soderblom et al. (1998b) speculate 
HD 180445 might be a spectroscopic binary. There is the hint of a second
pair of spectral lines which could be a tidally-locked secondary
causing the chromospheric emission. Therefore the age of HD 180445 is
in question, but Soderblom et al. (1998b) conclude HD 202917 is most
likely a single, very-active star with an age younger than the
Pleiades. Finally, 
Zuckerman \& Webb (2000) suggest HD 202917 is a member of the
co-moving cluster, Tucanae Association, with an age of $\sim$40 Myr.

HD 105 was identified in Favata et al. (1995) as having a high lithium
abundance (log N(Li)$=$3.4). Using a distance determined by the
Hipparcos mission, Favata et al. (1998) recomputed the X-ray
flux density measured with the Einstein satellite and find 
log L$_x=$29.2 erg s$^{-1}$,
which implies an age a little younger than Pleiades. For
reference, the solar X-ray luminosity varies, but is 
log L$_x$$\sim$27 erg s$^{-1}$.  Jeffries (1995) 
lists a vsin$i$$=$ 13 km s$^{-1}$ in his sample of active,
Li-rich stars.

Gl 311 (HD 72905) has a high level of chromospheric activity and
rotates rapidly. Soderblom et al. (1985) measure log (R'$_{HK}$)$=-$4.7, and Dorren \&
Guinan (1994) find a rotation period of 4.7 days, from which an age
of 300 Myr may be assigned (Kirkpatrick et al. 2001).  
Recently Gaidos, Henry, \& Henry (2000) brought these data 
together with a Li measure (log N(Li)=2.8), derived (U,V,W) space motions
($+$18.9,$+$12.1,-3) and assessed Gl 311 belonged to the UMa moving group 
($+$13,$+$1,-8). From all indicators, it is evident Gl 311 is most likely 0.3 Gyr.

HD 220140 (V368 Cep) has been identified as an a X-ray source with 
log L$_x=$30.5 erg s$^{-1}$ (Pravdo, White, \& Giomi 1985) and light
variations with a period of 2.75 days assumed to be due to spots have
been measured (Heckert et al. 1990). 
Chugainov, Lovkaya, \& Petrov (1991) find no radial velocity
variations over several nights, indicating a single star. They derive 
space motions of (-22,-28,-4) that are marginally consistent with the Pleiades
according to Eggen (1975) (-11,-25,-8). They derive a Li abundance of
log N(Li)$=$3.0 from their spectra, and based on these data 
give an age of 0.05 Gyr, but it is probably closer to the Pleiades age of
0.125 Gyr from the rotational velocity. Therefore we assign a range
from 50-125 Myr. 

It has been speculated that Gl 503.2 (HD 115043) 
is a member of the UMa co-moving group based on
lithium abundance, kinematics, and chromospheric activity. Soderblom
et al. (1993a) compared the abundances of Li in several clusters and
found Gl 503.2 to lie between the Hyades and Pleiades in 
abundances for G and K stars. The space velocities ($+$15,$+$3,-8)
(Rocha-Pinto \& Maciel 1998) are similar to the canonical UMa
motions. Measurements of Ca II H and K lines (log R'$_{HK}=-$4.43) and 
rotational velocity (vsin$i$= 9 km/s) were made by Soderblom \& Mayor (1993c).  
In a solar-like star, such activity is thought to be due to a
youthful stage of high magnetic activity. When compared to other stars
in Henry et al. (1996), the activity level places Gl 503.2
at less than 0.5 Gyr, and is listed with the most active stars in Soderblom
et al. (1998a). 
 
Gl 577 (HD 134319) has similar chromospheric emission in the Ca II H
\& K lines (log R'$_{HK}=-$4.33) to Pleiades stars (Henry et al. 1996), but 
its kinematic motions (-33,-15,-1) 
are more consistent with the Hyades (-40,-16,-3) cluster 
(Rocha-Pinto \& Maciel 1998). Messina, Guinan \& Lanza (1999) 
ascertained a rotation period of
4.448 days from photometric variations thought to be due to 
dark spots on the surface. They note this rotation
period is about half that of most Hyades members, but it does 
correlate well with UMa group members (Dorren \&
Guinan 1994). Therefore its age might be as young as 300 Myr, but
could be as old as 600 Myr.

HD 70573 and RE 1507$+$76 are both listed in Jeffries (1995) as 
active, Li-rich stars with vsin$i$ = 11 \& 14 km s$^{-1}$ respectively
and Li abundances above Pleiades values with Li (6708\AA) EW of 
149 \& 214 m\AA\ respectively. Jeffries (1995) derives space motions of
(-41,-27,-18) and (-14,-17,-11), respectively. The space motions of 
HD 70573 are inconsistent with those of Pleiades members, but from the
Li and rotation, we conclude that it is a little older than Pleiades
at 0.2$\pm$0.1 Gyr. RE 1507$+$76 motions are 
closer to that of Pleiades members than either those of
UMa or Hyades, consistent with the Li abundance and rotation. We
therefore assign an age of 0.1 Gyr.

HD 102982 was identified in Henry et al. (1996) as being a highly 
chromospherically active
star due to emission in the Ca II H \& K lines
(log R'$_{HK}=-$3.86) giving an age less than Pleiades. 
Mason et al. (1998) surveyed the most active
stars listed in Henry et al. (1996) for multiplicity and 
found no companions to HD 102982 within 3 mag between
0.035$-$1.08$\arcsec$ using speckle techniques. 
Soderblom et al. (1998b) found HD 102982 to be a spectroscopic
binary (SB2), and therefore the activity thought to result from youth might
instead be caused by the close, less than a few AU, companion. This star (b=9)
also slipped past our galactic latitude requirement (b$>$15) because
it was already on the observation calendar when the galactic latitude
cut was made.

The presence of the tidally-locaked companion 
to HD 102982 and HD 180445 places their ages in question. 
Our search with the coronagraph
(0.5$-$4$\arcsec$) covers $\sim$ 20$-$150AU at the distance of these
stars (42 pc), so we did not expect 
to image a tidally-locked companion, but rather search for a stable, lower-mass companion
farther out. 

\textit{\textbf{K-type stars}}

HD 1405 was identified to have chromospheric emission (Bidelman 1985) and
found to have a periodic variation (P=1.7 days) of almost 0.1 mag,
indicating rotation of cool spots (Griffin 1992). Pleiades members
typically have rotation periods of a couple days. The radial velocity
was found to remain constant and that is consistent with the lack of a
close companion. 
Fekel (1997) reports a vsin$i$$=$23km s$^{-1}$, 
which agrees with 21 km s$^{-1}$ of Griffin (1992). The rapid rotation and
chromospheric emission with lack of a change in radial velocity is
consistent with a youthful, single star at 0.1 Gyr. 
 
HD 17925 is of particular interest because it is only 10 pc from the
sun. Favata et al. (1995) measured log N(Li)=2.88 and vsin$i$$<$8km s$^{-1}$ in
their sample of X-ray active stars. Henry et al. (1996) measure
vsin$i$$=$6km s$^{-1}$, find the H$\alpha$ absorption partially filled in by
emission, but conclude that it could be a binary due to line-width
variations seen over the three nights they observed. However, Abbot,
Pomerance, \& Ambruster (1995) report photometric modulation in a period of 7
days, and a change in the light curve over a three week period
consistent with a single, active star. The star has a large
log R'$_{HK}$ ($=$-4.30; Rocha Pinto \& Maciel 1998) and Fekel (1997) measures
space motions (-14,-17,-11) close to that of the Pleiades, so we
assign an age of 0.1$-$0.2 Gyr.

Lk H$\alpha$ 264 was first identified as an emission line
star (Herbig \& Rao 1972) and later determined to be 
a classical T-Tauri star still
associated with the Lynds 1457 cloud at only 65 pc (Hobbs, Blitz, \&
Magnami 1986). Even though it is farther than 50 pc, it is likely a
few Myr old star with an H$\alpha$ equivalent width (EW) of 100m\AA\
(Gameiro et al. 1993). Using several lines of the spectrum, 
Gameiro et al. (1993) also measured a vsin$i$=22km s$^{-1}$\ consistent with a
young, late-type star.

HD 21703 was identified in the Einstein survey with a large X-ray flux
(log f$_x$/f$_{bol}$=-2.25; Stocke et al. 1991), and Favata et al. (1995)
measure an upper limit of log N(Li) $\leq$ 1.74 and vsin$i$$=$ 14km s$^{-1}$,
placing this mid-K type star near the Pleiades age from these
indicators.

Three more K-type stars were selected from Jeffries (1995) study
of high activity stars, correlating rotation and H$\alpha$ emission
with lithium abundance (Table \ref{jeffries}). At this effective temperature, the gap between
the Hyades lithium abundance and the Pleiades has widened
significantly (Favata et al. 1993), and these objects are near Pleiades age. 
RE2131$+$23 and RE 0723$+$20 also have measured space motions
(-4,-23,-13), and (-6,-27,-15) very close to those of the Pleiades, supporting
their young ages.

Henry, Fekel, \& Hall (1995) present results of photometric monitoring of 
HD 82443 deriving a period of 5.43 days, and confirm it is single 
with a constant radial
velocity. Fekel (1997) measured a vsin$i$=6.2 km s$^{-1}$, consistent with this
earlier measure. With a log(R'$_{HK}$)$=-$4.20 (Soderblom 1985) and space
motions of (-14,-24,-1) (Soderblom \& Clements 1987 ), we conclude this star
is probably a little older than Pleiades age. Gl 354.1B is a wide 
(65$\arcsec$), proper motion companion to HD 82443, so we conclude it
is of a similar age and add it to our program under M-type stars.

HD 82558 is a chromospherically active, single star. Most
main-sequence K-type stars are undetectable at radio frequencies, but
Gudel (1992) detected a 3.6cm flux of about 300 $\mu$Jy and attributed
it to gyrosynchrotron electrons in magnetic flux tubes in regions of
high chromospheric activity. This star was selected for their study
because of the X-ray emission (log L$_x=$29.1 erg s$^{-1}$) detected
in the Rosat All-Sky Survey. Jeffries (1995) lists vsin$i$=25km s$^{-1}$, 
Li EW $=$ 219 m\AA, and space motions of (-13,-5,-4) consistent
with an age of $<$ 100 Myr. 

Henry et al. (1995) found photometric variations in Gl 174 (HD 29697) 
with a period
of 3.9 days which they attribute to the presence of spots, deriving a
vsin$i$= 7km s$^{-1}$. They also measure a H$\alpha$ EW of 200 m\AA\ 
and Li EW of 79 m\AA. Using the conversion of Soderblom et al.
1993b, we convert this Li measure into an abundance of log N(Li) = 0.95. 
Eggen (1996) observed many low-mass stars and found photometric
variations in Gl 174, correlated that with Ca II
emission, and concluded that it was 60 Myr old. We conclude from the
lithium abundance and chromospheric emission that it is probably
closer to Pleiades age, or about 100 Myr old.

Gl 517 (HD 118100) has a high X-ray flux density (log
L$_x$/L$_{bol}$=-3.08; Sterzik, \& Schmitt 1997)
typical of young stars approaching the main-sequence, and Favata,
Micela, \& Sciortino (1997) measure log L$_{x}=$29.54 erg s$^{-1}$ which is
consistent. EUV activity and flares were observed from Gl 517
(Tsikoudi \& Kellett 1997). Favata et
al (1997) measure a Li (EW)$=$25m\AA\ and rapid rotation, concluding
this star is a BY Dra of a few tens of million years old.

Since Gl 879 (HD 216803) is the common proper motion companion 
to Fomalhaut, a
well-studied A-type star with a debris disk (Holland et al.
1998; Staplefeldt et al. 2004), determination of its age has been very important to
disk formation theories. Barrado y Navascues et al. (1997) 
studied Gl 879 to determine the age of both components placing several factors
together. They note the lithium measure of log N(Li)=0.6 makes the star
younger than the Hyades but older than Pleiades, while the X-ray
activity (logL$_{x}=$28.1; Favata et al. 1997) places the star 
closer to Hyades age. The rotation (vsin$i$$<$4km s$^{-1}$) also places the
star between the two clusters, but more consistent with the younger
Pleiades members, so they
conclude an age of 200$\pm$100 Myr for both Gl 879 and Fomalhaut.

Fekel (1997) finds HD 160934 has a constant radial velocity and
a high vsin$i$(=16.4km s$^{-1}$). ROSAT detected this star with log
f$_x$/f$_v=-2.07$ (Schachter et al. 1996) and Henry et al. (1995)
measured photometric variability due to spots with a period of $\sim$2
days. Therfore, from the rotation and chromospheric X-ray emission, 
it is most likely a Pleiades age star. 

HD 177996 was measured to have considerable Ca II emission
(log R'$_{HK}$=-4.17; Henry et al. 1996), and Soderblom et al. (1998a)
derive logN(Li)=1.04, placing the star's age between Hyades and
Pleiades. They do find it to be a double-lined spectroscopic binary, 
but detect lithium in both stars and suggest an intermediate age of
0.5 Gyr. 

Jeffries (1995) lists HD 197890 as a very active star with a
vsin$i$=170 km s$^{-1}$ and log N(Li)=3.1 (Anders et al. 1993). They derive space
motions of (-6,-13,$+$1), similar to the Pleiades. (Favata et al. 1998)
measure an X ray activity (log L$_{x}$=30.2), consistent with a 100
Myr star.

\begin{figure}
\figurenum{A1}
\epsscale{0.6}
\plotone{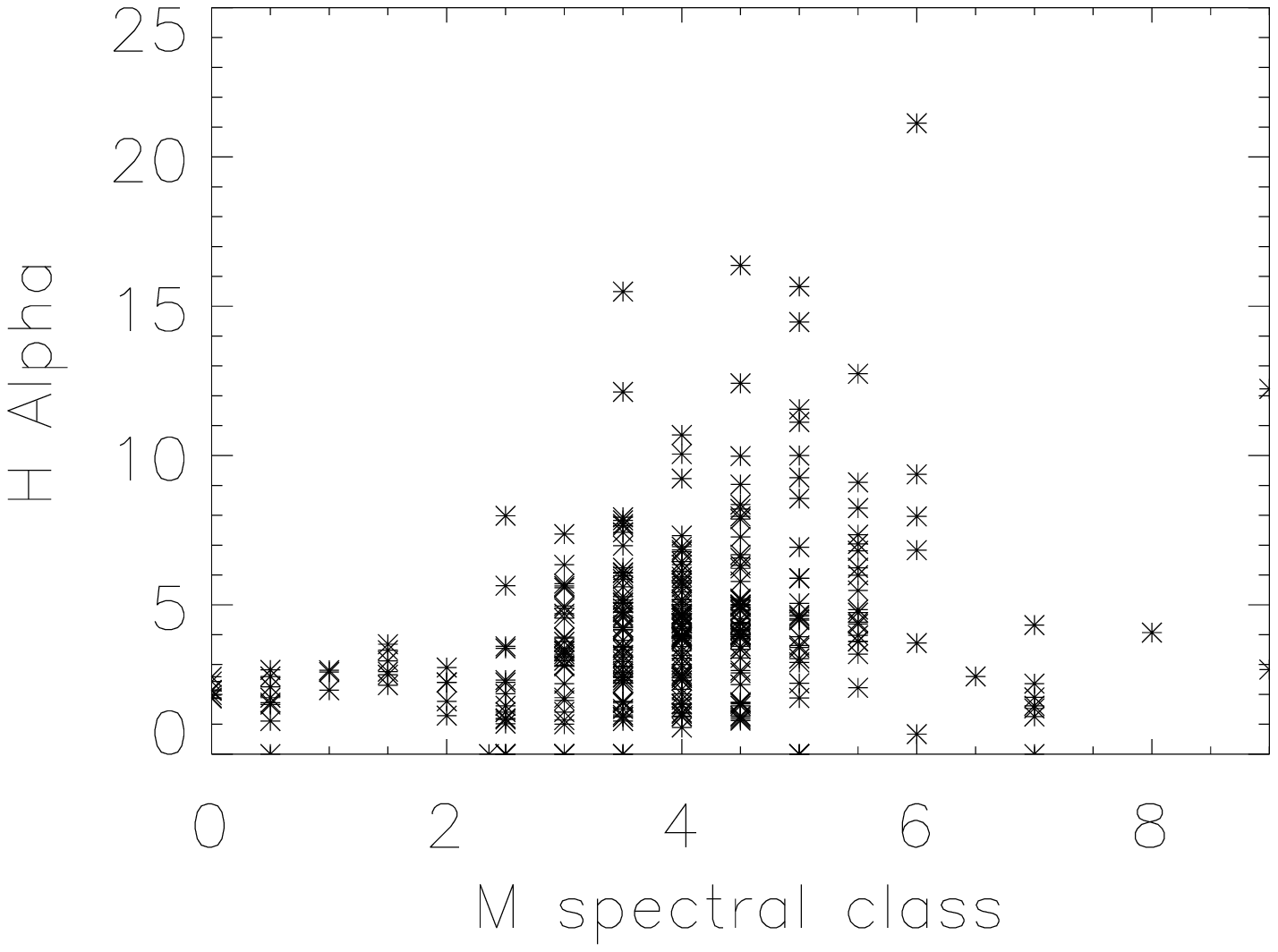}\label{mstars}
\figcaption{The H$\alpha$ measurement plotted against spectral type for
those M-type stars with a measurement in Reid, Hawley \& Gizis
(1995). We took single targets from the top 10\% of this distribution across
all M spectral classes.}
\end{figure}

\textit{\textbf{M-type stars}} Usually it is a little 
harder to determine an age for single M-type stars. For example, lithium
is destroyed within a few million years in the fully convective
interiors and cooler surfaces tend to display a smaller amount of
chromospheric activity. We list eight stars selected as the most 
active, single M-type stars from Reid, Hawley \& Gizis (1995)
in Table \ref{halpha}  and with higher than normal H$\alpha$ activity, 
thought to be Pleiades age (Figure \ref{mstars}). The H$\alpha$ EW
ranges from 2\AA\ at M0 to 9\AA\ at M5 for the top 10\% of Hyades members
(Reid, Hawley \& Mateo 1995), and
from 4\AA\ at M0 to 12\AA\ at M5 for the same tier of Pleiades members
(Hodgkin, Jameson, \& Steele 1995). Our
targets are most consistent with the top members of the Pleiades
cluster, suggesting ages of 100 Myr. 
Some M stars ($<$10\%) have been observed to flare, and 
any of these measures listed in Table \ref{halpha} could 
have been taken during a flare. Therefore we suggest one should be cautious 
in determining ages from one observational criterion.

Gl 875.1 was detected in EUV flare activity (Tsikoudi \& Kellett 1997)
and Mathioudakis et al. (1995) measure a rotation period of 1.64 days
and H$\alpha$ EW of 3.9\AA. The rotation period is indicative of
Pleiades age stars, but the H$\alpha$ is more consistent with an older
star, perhaps UMa. Therefore we assign an age of 0.2 Gyr.

\textit{\textbf{TW Hydrae Association}}
Over the last few years mounting evidence has suggested that a number
of young, active stars in the vicinity of TW Hydrae form a physical
association with an age of $\sim$ 10 Myr (Kastner et al. 1997; 
Webb et al. 1999; Soderblom et al. 1998c). 
At an approximate distance of 60 pc, 
the ``TW Hydrae Association'' (TWA) is a region of 
recent star formation nearest to the Sun (Kastner et al. 1997). 
Webb et al. (1999) added HR 4796A and 
identified five new systems (seven members), in which each 
system is characterized by the presence of X-ray emission, 
H$\alpha$ emission, and strong lithium absorption associated with young stars.
The currently identified eleven
systems are shown to have similar space motions implying physical association and
a possible common origin (Webb et al. 1999). We surveyed 6 of the members,
including TWA 1, 5, 6, 7, 8B, and 10 for possible brown dwarf
companions. The multiple stars HR 4796A and HD 98800A/B were observed
in a sister program for circumstellar disks.

\newpage 

\begin{deluxetable}{llccccccc}
\tablenum{1}
\tablecaption{Survey Stars}
\tabletypesize{\scriptsize} 
\tablehead{\colhead{Primary} & \colhead{other names} &\colhead{$\alpha$ (J2000)} &
\colhead{ $\delta$ (J2000)} & 
\colhead{m$_{H}$} & \colhead{Spec type} & \colhead{D (pc)} &\colhead{b} &\colhead{Approx age (Gyr)} }  
\startdata
\hline
HR 7329 & HD 181296 &19 22 51.23 & $-$54 25 26.2 & 5.03 & A0V &  47.7* & $-$26 &  0.01$-$0.04 \\
HR 8799 &HD 218396 &23 7 28.72  &  $+$21 08 03.4& 5.33 & A5V & 39.9* & $-$36  &  0.01$-$0.04  \\
SAO 170610  &HD 37484  &05 37 39.63  & $-$28 37 34.6 & 6.38  & F3V  &  59.5*  & $+$35 &  0.02 - 0.1 \\
HD 35850 & &05 27 04.77 & $-$11 54 03.5& 5.19 & F7V &  26.8* & $-$24 & 0.1  \\
HD 209253 & &22 02 32.98 & $-$32 08 01.6  &5.53 & F6/F7V & 30.1* &$-$53 &  0.2--0.3  \\
HD 105  & &00 05 52.56 & $-$41 45 11.5  & 6.17  & G0V  & 40.2*  & $-$73 & 0.1  \\
HD 70573 & &08 22 49.94  & $+$01 51 33.4 &  7.36 & G6V & 32  & $+$21  & 0.2--0.3 \\
GL 311 &HD 72905  &08 39 11.73  & $+$65 01 15.2& 4.25 & G1V & 14.3* & $+$36 & 0.3  \\
GL 503.2 &HD 115043&13 13 37.10 & $+$56 42 30.1 & 5.45 & G2V & 25.7*  & $+$60 & 0.3  \\
HD 102982 && 11 51 09.14 &$-$51 52 32.3& 6.96& G3V & $\sim$42 &$+$10 & 0.1 $?$   \\
GL 577 &HD 134319 &15 05 50.16 & $+$64 02 49.8 & 6.88 & G5V &  44.3*  & $+$47 & 0.3-0.6   \\
HD 135363 & RE 1507$+$76 &15 07 56.24 & $+$76 12 02.4 & 7.19  & G5V  & 29.4*  & $+$38 &  0.1   \\
HD 180445  & & 19 18 12.65 & $-$38 23 04.6 & 6.84 & G8V & 41.7*  & $-$21 & 0.20$?$   \\
HD 202917 &&21 20 49.95  & $-$53 02 03.0 & 7.13 & G5V & 45.9* & $-$43 & 0.04-0.1   \\
HD 220140 & V368 Cep&23 19 26.56  & $+$79 00 12.4& 5.87 & G9V & 19.7* & $+17$ &0.05--0.1  \\
HD 1405 &PW And  & 00 18 20.76  & $+$30 57 22.0 & 7.35 & K2V & 47.9 & $-$31  & 0.1   \\
RE 0041$+$34 &QT And & 00 41 17.26 & $+$34 25 17.7 &  6.41 & K7V & 15.0 & $-$28 & 0.03   \\
HD 17925&EP Eri  &02 52 32.15 & $-$12 46 11.1 & 4.22& K1V& 10.4* & $-$58&  0.1--0.2  \\
LkHa 264 &WY Ari& 02 56 37.65  & $+$20 05 36.0 & 10.14 & K3V & $\sim$65 & $-$34&  0.01  \\
HD 21703 &AK For &03 29 22.88 & $-$24 06 03.1 &  6.59& K4V& 31.7*& $-$54  & 0.1  \\
GL 174 &  HD 29697 &04 41 18.82  & $+$20 54 05.5 & 5.75 & K3V & 13.5*  & $-$16 & 0.06  \\
G 88-24&RE 0723$+$20 &07 23 43.68 & $+$20 25 02.5& 7.19 & K5V & 23.0  &$+$16 & 0.1   \\
HD 82443 & GJ 354.1A&09 32 43.78 & $+$26 59 18.5& 5.28 & K0V & 17.8*  & $+$46 & 0.2  \\
HD 82558 & GJ 355 &09 32 25.87  & $-$11 11 04.8& 6.03 & K0V & 18.3* &$+$28 & $<$0.1   \\
TWA 6      & &10 18 28.86 & $-$31 50 03.3& 6.89 & K7V & $\sim$60  & $+$21 & 0.01  \\
GL 517 & HD 118100& 13 34 43.19 & $-$08 20 31.3 & 6.46 &  K5V &  19.8*  &$+$53 &  0.05   \\
HD 160934 &    &17 38 39.81 & $+$61 14 14.0&   7.21 & K8V & 24.5*  &$+$32 &0.1   \\
HD 177996  &  SAO 229520 &19 08 50.45 & $-$42 25 41.5 & 5.97 & K1V & 31.8*  &$-$21 &   0.5   \\
HD 197890 & BO Mic& 20 47 44.97 & $-$36 35 40.7 & 7.68 & K0V & 44.4*  &$-$38  & 0.1  \\
G 145-43 &RE 2131$+$23 &21 31 01.50 & $+$23 20 06.2 & 6.16 & K5V & 25.1* &$-$20  & 0.1    \\
GL 879 & HD 216803  &22 56 24.08 & $-$31 33 56.1 & 3.78 & K4V & 7.6* &$-$64  &0.2  \\
PS 176	  &   &01 19 27.34 & $-$26 21 55.3 &  8.6   &M3 & 25.5  & $-$83 &0.1   \\
GL 207.1  &  G 99-17  & 05 33 45.12 & $+$01 56 47.0 & 7.44  &M2.5& 14.6 &$-$16  & 0.1  \\
LH 2026  & LP 605-23  & 08 32 30.28 & $-$01 34 31.1& 11.48   &M6& 19.7 & $+$21  &0.1   \\
GL 354.1B &&09 32 48.48 & $+$26 59 44.7& 8.86 & M5.5V & $\sim$18  & $+$46 & 0.2 \\
TWA 7     & & 10 42 30.36 & $-$33 40 17.9 & 7.44 &  M4V & $\sim$60  &$+$22 & 0.01  \\
LH 2320  & G 44-43 & 10 52 15.09 & $+$05 55 10.0 & 8.21   &M5  & 9.9 & $+$55 & 0.1  \\
LP 263$-$64 & & 11 03 10.21 & $+$36 39 07.3 & 9.03   &M3.5& 23.3  & $+$65 & 0.1  \\
Steph 932 & &11 15 54.37  & $$+$$55 19 51.0 & 6.49 & M0.5V & 15.0 & $+$57 &0.1  \\
TWA 5 & CD $-$33 7795&11 31 55.40 & $-$34 36 27.3 & 7.35 & M3V & $\sim$60 & $+$25 & 0.01  \\ 
TWA 8B  & & 11 32 41.37 & $-$26 52 08.8 &  9.38  & M3 &$\sim$60 & $+$32 &0.01    \\
TWA 10  & &12 35 04.48 & $-$41 36 39.9 & 8.43   & M3 & $\sim$60 & $+$21  &0.01    \\
LP 390$-$16 &  & 18 13 06.37 & $+$26 01 51.8 & 9.10  &M4& 16.6 & $+$19 & 0.1   \\
GL 875.1 &LH 3861 & 22 51 52.87 & $+$31 45 16.6 &  7.42 & M3.5V &  14.2*  & $-$25 & 0.3   \\
GJ 1285	  & AF Psc   &23 31 44.81 & $-$02 44 39.7& 10.85   &M4& 29.2 & $-$59 & 0.1  \\
\hline
\enddata
\tablenotetext{*}{HIPPARCOS determined distances.}
\tablenotetext{a}{TWA star distance are based
on cluster membership, and otherwise, distance are photometric.}
\tablenotetext{b}{This table was constructed through extensive use of the SIMBAD database.}
\label{sample}
\end{deluxetable}

\newpage

\begin{deluxetable}{lccccccccc}
\tablenum{2}
\tablecaption{Summary of Point-sources Candidates}
\tabletypesize{\scriptsize} 
\tablehead{
\colhead{} & \colhead{H$_{prim}$}   & \colhead{Spec type}   & \colhead{D }   & 
\colhead{Age} & \colhead{m$_H$}  & \colhead{Sep} & \colhead{PA}& \colhead{cpn/}\\
\colhead{} & \colhead{mag}   & \colhead{(prim)}   & \colhead{(pc) }   & 
\colhead{(Gyrs)} & \colhead{}  & \colhead{$\arcsec$} & \colhead{}& \colhead{bgd?}}
\startdata
HD102982 &  6.9& G3V & ~42 & $<$ 0.30  & 10.9 &  0.9$\arcsec$  & 28.6 & cpn  \\
LP 390-16   &  9.10  &M4& 16.6    &0.1 &  14.4 &  1.45$\arcsec$ & 226.0 & bgd\\	
GL503.2 & 5.45 & G2V & 25.7 & 0.3 &  10.5 &  1.55$\arcsec$& 354.9 & cpn \\
TWA 5      & 7.35 & M3V & ~55 & 0.01 &   12.2 & 1.96$\arcsec$& 358.8 & cpn \\
TWA 6      &  6.9 & K7V & ~55 & 0.01 &   19.9 & 2.54$\arcsec$& 278.7 & bgd\\
TWA 7      & 7.4 &  M4V & ~55 &  0.01 &  16.8 & 2.47$\arcsec$& 142.2 & bgd \\
\hline
HR7329 & 5.03 & A0V &  47.7 &  0.01--0.04 &  11.9 & 4.17$\arcsec$ & 166.8 & cpn\\
GL577 & 6.88 & G5V &  44.3 &  0.3 &  10.9  &  5.34$\arcsec$& 260.6 & cpn \\
HD220140 & 5.87 & G9V & 19.7 & 0.05 &  7.8 &  10.85$\arcsec$ & 216.3 & cpn \\
GL875.1 & 7.4 & M3.5V &  14.2 & $<$ 0.3 &  20.2 &  5.42$\arcsec$& 250.4 & bgd \\
HD160934 &    7.21 & K8V & 24.5 & 0.1 &  16.4 &  8.69$\arcsec$& 234.8 & cpn \\
HD180445  & 6.84 & G8V & 41.7 & $<$ 0.20 &   9.9  &  9.45$\arcsec$ & 52.5 & cpn\\ 
          &      &     &      &          &   20.8 &  2.88$\arcsec$ & 337.0 & \\ 
 	&      &   &     &  &  18.7   &4.51$\arcsec$ & 108.3 & \\
 	&      &   &     &  &  17.8   & 13.87$\arcsec$ & 275.6 & \\
 	&      &	   &     &  &  21.4  & 11.61$\arcsec$ &  200.7 & \\
LHS 2320    & 8.2   &M5  & 9.9    &0.1 & 13.5  & 14.92$\arcsec$ & 96.9 & bgd \\	
 	&      &   &     &  &  15.9   & 15.13$\arcsec$ & 93.6 & \\
RE0723+20 & 7.19 & K5V & 23.0 &   0.2   &  14.6 &  5.12$\arcsec$  & 322.7 & bgd \\
 	&      &	   &     &  & 17.7   & 9.78$\arcsec$ & 277.4 & \\
 	&      &	   &     &  &  17.8    &  13.85$\arcsec$ & 326.9 & \\
HD177996  & 5.97 & K1V & 31.8 &  0.5 &  19.6 &  5.20$\arcsec$ & 300.9   &bgd \\
 	&      &	   &     &  & 15.6  & 8.08$\arcsec$ & 301.1 & \\
	&      &   &     &  & 17.9     & 12.50$\arcsec$ & 244.2 & \\
GL 207.1    &  7.4  &M2.5& 14.6    &0.1 & 18.0, 17.6 &  4.1.4.2$\arcsec$ & 197.4,199.1& \\
	    &      &	   &     &  &   17.7  & 9.7$\arcsec$ & 289.6 &  \\
HD82443 & 5.28 & K0V & 17.8 &   0.2  &  16.9 &  6.86$\arcsec$& 195.3 & \\
HR8799 & 5.33 & A5V & 39.9 & 0.01--0.04  &  21.6 &  13.71$\arcsec$ & 15.9 & \\
       &      &   &     &  & 20.4   & 15.68$\arcsec$ & 114.3 & \\
RE1507+76  & 7.19  & G5V  & 29.4  & 0.1   &  16.6 &  17.11$\arcsec$& 8.3 & \\
HD202917 & 7.13 & G5V & 45.9 &  0.1  &   18.1 &  13.62$\arcsec$& 230.1 & \\
HD17925&  4.22& K1V& 10.4& 0.1 &  17.47 &  15.36$\arcsec$ & 294.6  &\\
LkHa 264 & 10.14 & K3V & ~65 &   0.01  & 16.8 &  9.28$\arcsec$  & 220.4  &\\ 
 	&      &   &     &  &  21.0   & 9.51$\arcsec$ & 233.4 & \\
GL174 & 5.75 & K3V & 13.5 & 0.06  &  19.4 &  14.43$\arcsec$& 199.2 & \\
HD197890 & 7.68 & K0V & 44.4 &  $<$ 0.1 &  17.7 &  9.04$\arcsec$ & 20.4 & \\
GL354.1B & 8.86 & M5.5V & ~18 & 0.2 &  16.51 &  9.51$\arcsec$ & 317.0 & \\ 
GJ 1285	    & 10.85   &M4& 29.2    & 0.1 &  14.8 &  19.04$\arcsec$& 50.4 & \\
\enddata
\tablenotetext{1}{If a survey star is not found in this list, then there 
were no point-sources detected in the NICMOS images}
\tablenotetext{2}{All separations have errors of 0.08$\arcsec$}
\label{cand}
\end{deluxetable}

\clearpage

\newpage

\begin{deluxetable}{lccc}
\tablenum{3}
\tablecaption{Proper Motion of Primary Stars w/ Followup}
\tablehead{
\colhead{Star}  &   \colhead{$\mu_{\alpha}$ (mas/yr)} & \colhead{$\mu_{\delta}$ (mas/yr)}  & \colhead{Reference}}
\startdata
HD 102982 & -0.057     &  -0.081  & Hog et al (2000)	\\ 
Gl 503.2  & 0.112      & -0.018  &   Perryman et 1997 \\
Gl 577    & -0.121     &   0.112  & Perryman et al 1997 \\
HD 180445&	0.099	&  -0.093   	& Perryman et al (1997)	\\
HD 220140&	0.201	&  0.072   	& Perryman et al (1997)	 \\
HD 160934&     -0.031  &  0.059		& Perryman et al (1997)	 \\
LP 390-16&	0.207	&  -0.036 	& Luyten (1979)	\\
HD 177996&	0.023	&  -0.120	& Perryman et al (1997)	\\
Re 0723+20&	-0.055	&  -0.238	& Montes et al (2001)	\\
LHS 2320&	-0.673    &  -0.094	& Bakos, Sahu,\& Nemeth (2002) \\
Gl 875.1&       0.527	&  -0.050	& Perryman et al (1997)\\
TWA 7	&	-0.122	&  -0.029	& Hog et al (2000)	\\
TWA 6	&	-0.060   & -0.020	& Webb et al (1997)	\\
\enddata
\label{pmA}
\end{deluxetable}

\clearpage

\clearpage

\clearpage

\clearpage

\clearpage

\clearpage

\clearpage
\begin{deluxetable}{lccccc}
\tablenum{A1}
\tablecaption{Stars from Jeffries (1995)}
\tablehead{
\colhead{Star}  &  \colhead{Sp Type} & \colhead{vsin$i$ } &  
\colhead{H$\alpha$ (EW)}& \colhead{Li(EW)} & \colhead{log N(Li)$^1$} \\
      &		 &  \colhead{km s$^{-1}$}	 &   \colhead{(m\AA)} & \colhead{(m\AA)} &   \\
}
\startdata
RE 0041$+$34 &  K7   & 15       &  300   & 127 & 1.3\\
RE 0723$+$20 &  K5   & 12	&  350   & 105 &1.2\\
RE 2131$+$23 &  K5   & 70	&  517   & 215 & 1.9\\
\enddata
\tablenotetext{1}{Derived from EW Li using Soderblom et al. (1993b)}
\label{jeffries}
\end{deluxetable}

\begin{deluxetable}{lcc}
\tablenum{A2}
\tablecaption{Active M Stars from Reid, Hawley \& Gizis (1995)}
\tablehead{
\colhead{Star } &  \colhead{Sp Type}&   \colhead{H$\alpha$ (\AA)}  \\
}
\startdata
Steph 932 &  M 0.5 & 2.82\\
Gl 207.1   &   M 2.5& 7.98\\
PS176      &   M 3 & 7.37\\
LP263-64  &   M 3.5 & 7.83\\
LP390-16  & M 4    & 10.05\\
GJ 1285    &   M 4.5 & 16.37\\
LHS2320    & M 5   & 14.47\\
LHS2026    &  M 6   & 21.13 \\
\enddata
\label{halpha}
\end{deluxetable}

\newpage

\end{document}